\documentclass[11pt,reqno]{amsart}
\usepackage{amssymb}

\usepackage{enumitem}
\usepackage{amsthm,amsfonts,amssymb,euscript,color,bbm}
\usepackage{comment}
\usepackage{mathtools}
\usepackage{stmaryrd}
\usepackage{mathrsfs}
\usepackage{pifont}
\usepackage{amsfonts,amsmath,amssymb,amsthm,amscd,cancel}
\usepackage{graphicx}
\usepackage[]{geometry}
\usepackage{verbatim}
\usepackage{mathrsfs}
\usepackage{fancyhdr}
\usepackage{footnpag,footmisc}
\usepackage[normalem]{ulem}
\usepackage{float}

\usepackage{cite}
\usepackage{hyperref}
\usepackage{engord}
\usepackage[displaymath]{lineno}
\usepackage[utf8]{inputenc}

\theoremstyle{definition}
\newtheorem{theorem}{Theorem}[section]
\newtheorem{lemma}{Lemma}[section]
\newtheorem{proposition}{Proposition}[section]
\newtheorem{corollary}{Corollary}[section]
\newtheorem{definition}{Definition}[section]

\newtheorem{remark}{Remark}[section]

\newtheorem*{mth}{Main Theorem}

\allowdisplaybreaks[4]

\DeclareMathAlphabet{\mathsfsl}{OT1}{cmss}{m}{sl}
\numberwithin{equation}{section}

\newcommand{\D}{\mathrm{d}}

\def\nablas{\mbox{$\nabla \mkern -13mu /$ }}
\def\Deltas{\mbox{$\Delta \mkern -13mu /$ }}

\def\divs{\mbox{$\mathrm{div} \mkern -13mu /$ }}

\begin{document}
	
	\title[]{The uniqueness of smooth stationary black holes}

 \author{Yishen Huang}
 	\address{Department of Mathematics, Sun Yat-sen University, Guangzhou, China}
	\email{huangysh36@mail2.sysu.edu.cn}
 
	\author{Jiongyue Li}
	\address{Department of Mathematics, Sun Yat-sen University, Guangzhou, China}
	\email{lijiongyue@mail.sysu.edu.cn}
 
 \author{Junbin Li}
 	\address{Department of Mathematics, Sun Yat-sen University, Guangzhou, China}
	\email{lijunbin@mail.sysu.edu.cn}
 \author{Xi-Ping Zhu}
 	\address{Department of Mathematics, Sun Yat-sen University, Guangzhou, China}
	\email{stszxp@mail.sysu.edu.cn}
	
	\maketitle

\begin{abstract}
We show that the domain of outer communications of a smooth, asymptotically flat,  stationary  vacuum spacetime describing a single rotating non-degenerate black hole is isometric to the domain of outer communications of a subextremal Kerr solution under reasonable geometric and physical assumptions, without analyticity or any smallness assumptions. The method is to apply Harrison transform  to  construct a family of solutions of Einstein--Maxwell equations. Combining null area comparison with the area-charge inequality, we show that, in the limiting transformed geometry, the bifurcate sphere has area radius one and constant electric and magnetic flux densities, with the original ADM mass normalized to $M=1$. These identities imply the technical assumptions of Ionescu--Klainerman, yielding the identification with Kerr.
\end{abstract}

\section{Introduction}\label{sec:introduction}

\subsection{Previous results}
The black hole uniqueness conjecture is a long-standing problem in general
relativity. It asserts that the \emph{domain of outer
communications} of a regular, four-dimensional, asymptotically flat, stationary vacuum black hole is isometric to the exterior of a Kerr black hole. The domain of outer communications refers to the region which is causally connected to faraway region. A physical motivation for the conjecture is the expectation that an isolated dynamical black hole will approach
a stationary state at late times due to its persistent emission of gravitational radiation. Thus the conjecture, if true, would characterize all possible asymptotically states of the general vacuum evolution.

The vacuum Einstein equations and the stationarity condition take the form
\begin{equation*}
\operatorname{Ric}(\mathbf g)=0,\quad \mathcal L_{\mathbf T}\mathbf g=0 ,
\end{equation*}
on a four-dimensional Lorentzian manifold, where $\mathbf T$ is the stationary Killing field, future timelike near
infinity and normalized by $\mathbf g(\mathbf T,\mathbf T)\to-1$ there.
In Boyer--Lindquist coordinates $(t,r,\theta,\phi)$, the Kerr metric first found in \cite{Kerr1963}, describing the mass $M>0$ and angular momentum $J=Ma$ is
$$
 \mathbf g_{\rm Kerr}=
 -\frac{\Delta}{\rho^2}\bigl(dt-a\sin^2\theta\,d\phi\bigr)^2
 +\frac{\sin^2\theta}{\rho^2}\bigl((r^2+a^2)d\phi-a\,dt\bigr)^2
 +\frac{\rho^2}{\Delta}\,dr^2+\rho^2\,d\theta^2 ,
$$
where
$$
 \rho^2=r^2+a^2\cos^2\theta,\qquad \Delta=r^2-2Mr+a^2 .
$$
We say a Kerr spacetime is subextremal if $|a|<M$, and extremal if $|a|=M$. In these
coordinates the Kerr exterior is the region $r>r_+$, where
$r_+=M+\sqrt{M^2-a^2}$ is the location of the boundary of the black hole, called the event horizon. The member with $a=0$ is the Schwarzschild solution. 

In this paper, we prove the uniqueness of subextremal Kerr solution in stationary black hole solution, we first state our main result in a rough version.
\begin{mth}[rough version]
Under certain geometric and physical assumptions, the domain of outer communications of a smooth, asymptotically flat, stationary vacuum spacetime describing a single non-degenerate rotating black hole is  isometric to the domain of outer communications of a subextremal Kerr spacetime with $0<|J|<M^2$.
\end{mth}

We call a stationary black hole, or precisely a horizon $\mathcal{H}$ is non-rotating if $\mathbf{T}$ is tangent to its null generators, and rotating if this tangency fails somewhere, which is introduced by Chru\'sciel--Costa~\cite{ChruscielCosta2008}. In vacuum, if a stationary black hole is non-rotating, the it was shown \cite{ChruscielCosta2008,SudarskyWald1993,Israel1967,BuntingMasoodulAlam1987} that its exterior (i.e., domain of outer communications) is isometric to the exterior of a Schwarzschild solution. In rotating case, Hawking \cite{Hawking1972} shows that there is another non-vanishing vector field $\mathbf{K}$ on the horizon $\mathcal{H}$ tangent to the null generators of the horizon, which is Killing to any order along $\mathcal{H}$.  Then along $\mathcal{H}$, 
$$\nabla_\mathbf{K}\mathbf{K}=\kappa\mathbf{K}$$
for some constant $\kappa\ge0$, which is called the surface gravity. When $\kappa>0$, the black hole is called non-degenerate and otherwise degenerate. 

We first focus on the non-degenerate case. In real analytic case, Hawking's rigidity theorem \cite{Hawking1972} asserts $\mathbf{K}$ can be extended to the whole exterior which makes the exterior axisymmetric. Carter--Bunting \cite{Carter1971,Robinson1975} proved that the exterior of a stationary and axisymmetric asymptotically flat black hole  must necessarily be Kerr exterior. This latter step does not require analyticity. See Chru\'sciel--Costa~\cite{ChruscielCosta2008} for a more complete and rigorous treatment.

However, the analyticity is hard to justify, there is no reason to expect that an evolution solution  eventually becomes analytic. So it is natural to consider the uniqueness problem without analyticity. One may refer to a nice survey of Ionescu--Klainerman \cite{IonescuKlainerman2015}. A first major advance in smooth category is the work of Ionescu--Klainerman~\cite{IonescuKlainerman2009}, in which the uniqueness of Kerr is settled under the assumption of an identity relating the Ernst potential and the Killing scalar on the bifurcate sphere, by using  Mars's curvature characterization of Kerr
\cite{Mars1999}. Our proof is to show that this identity is always true for a smooth, asymptotically flat, stationary vacuum spacetime with a single non-degenerate rotating black hole. 
The bifurcate sphere is the intersection of the future and past horizons $\mathcal H^+$ and $\mathcal H^-$. The existence of a bifurcate sphere fwas shown by Racz--Wald \cite{RaczWald1992,RaczWald1996} under certain global assumptions. In fact, Ionescu--Klainerman~\cite{IonescuKlainerman2013} showed that a local extension without additional symmetries (and hence not isometric to Kerr) can be extended in a neighborhood of the horizon away from the bifurcate sphere. It is then reasonable to assume the existence of the bifurcate sphere. One may ask whether one can go back to Hawking's strategy to show that the Hawking Killing field $\mathbf K$ can be extended to the whole exterior without analytic assumption and then apply Carter--Robinson's result.  In \cite{AlexakisIonescuKlainerman2010a}, Alexakis--Ionescu--Klainerman  showed that the Hawking Killing field can always be extended to a full neighborhood of the bifurcate sphere. In their later works \cite{AlexakisIonescuKlainerman2014,AlexakisIonescuKlainerman2010b}, the Hawking Killing field was shown to be extended under certain smallness assumptions on the bifurcate sphere or on a Cauchy hypersurface.  The basic tool of these works is the Carleman type estimates, allowing them to prove uniqueness for the ill-posed problem.  The main obstruction is some convexity conditions (see \cite{IonescuKlainerman2015}), which is not known for a general stationary black hole allowing ergoregion in which $\mathbf{T}$ is spacelike. Very recently, Radu  \cite{Radu2026a} gave a proof of the uniqueness under a nontrapping condition for null geodesics orthogonal to $\mathbf{T}$, answering a conjecture proposed in \cite{IonescuKlainerman2015}.  In another paper \cite{Radu2026b}, Radu was able to prove uniqueness under the assumption that the  ADM angular momentum is small enough, and certain geometric bounds are also assumed. In \cite{Radu2026b}, the author introduced a powerful frame work, the Harrison transformation \cite{Harrison1968}, to construct solutions of Einstein--Maxwell equations from the original stationary vacuum metric. Certain rigidity can be deduced using null area comparison by Raychaudhuri equations. Our proof of this paper will be  based on this framework. 

We also mention the uniqueness results for stationary electrovacuum black hole in analytic case \cite{Mazur1982,Bunting1983,ChruscielCostaHeusler2012} and smooth case \cite{Yu2010,LaiLiYu2023}.

\subsection{The precise statement}

We now give the precise statement. The assumptions are basically from \cite{IonescuKlainerman2015} with minor modifications.

\begin{definition} 
Let $(\mathbf M,\mathbf g)$ be a smooth  Lorentzian $4$-dimensional manifold 
with a smooth stationary Killing field $\mathbf T$ such that
$$
 \operatorname{Ric}(\mathbf g)=0,\quad
 \mathcal L_{\mathbf T}\mathbf g=0.
$$
We call $(\mathbf{M},\mathbf{g})$ a regular non-degenerate stationary vacuum spacetime if the following conditions are verified.
\begin{enumerate}[wide, font=\bfseries, leftmargin=*, labelsep=0.5em]
\item[(1)] (\emph{Asymptotic flatness.})
There is an open subset $\mathbf M^{(end)}\subset\mathbf M$
diffeomorphic to
$\mathbb R\times\{x\in\mathbb R^3:|x|>R_0\}$ for some sufficiently
large $R_0$. In the corresponding coordinates $(t,x^i)$, with
$r=|x|$, the metric satisfies
\begin{equation} \label{eq:af}
 \mathbf g_{00}=-1+\frac{2M}{r}+O_6(r^{-2}),\quad
 \mathbf g_{ij}=\delta_{ij}+O_6(r^{-1}),\quad
 \mathbf g_{0i}=O_6(r^{-2}).
\end{equation}
$u=O_m(r^{-b})$ means that $\partial_x^\alpha u=O(r^{-b-|\alpha|})$
for $|\alpha|\le m$. We assume $\mathbf T=\partial_t$, with $t=x^0$. We define the domain of outer communications by
$$
 \mathbf E=I^-(\mathbf M^{(end)})\cap I^+(\mathbf M^{(end)}).
$$

\item[(2)] (\emph{Completeness.})
The region $\mathbf E$ is globally hyperbolic, and every orbit of
$\mathbf T$ in $\mathbf E$ is complete and intersects a given
spacelike hypersurface $\Sigma_0$, whose portion in $\mathbf E$ is
Cauchy. We assume that $\Sigma_0$ is diffeomorphic to
$\{x\in\mathbb R^3:|x|>1/2\}$ and agrees with $t=0$ in
$\mathbf M^{(end)}$.

\item[(3)] (\emph{Smooth bifurcation sphere.})
Let
$$
 S=\partial I^-(\mathbf M^{(end)})\cap\partial I^+(\mathbf M^{(end)}).
$$
We assume that $S\subset\Sigma_0$ is a smooth embedded two-sphere
corresponding to $|x|=1$ under the identification of $\Sigma_0$ with
$\{|x|>1/2\}$. There is a neighborhood $\mathcal O$ of $S$ in
$\mathbf M$ such that
$$
 \mathcal H^+=\mathcal O\cap\partial I^-(\mathbf M^{ end}),\qquad
 \mathcal H^-=\mathcal O\cap\partial I^+(\mathbf M^{ end})
$$
are smooth embedded hypersurfaces diffeomorphic to $S\times(-1,1)$.
They are null and non-expanding, meaning that their null expansions
vanish, and they intersect transversely in $S$.

\item[(4)] (\emph{Tangency at the horizon.})
The stationary Killing field $\mathbf T$ is tangent to both
$\mathcal H^+$ and $\mathcal H^-$.
\end{enumerate}
\end{definition}

To state our theorem, we write
$$F_{\alpha\beta}=\nabla_\alpha\mathbf T_\beta, \quad \mathcal F_{\alpha\beta}=F_{\alpha\beta}
       +\frac{\mathrm i}{2}\epsilon_{\alpha\beta\mu\nu}F^{\mu\nu}.$$
By direct computation, $2\mathbf T^\beta\mathcal{F}_{\beta\alpha}$ is closed, there is a complex-valued function, called Ernst potential $\sigma$ so that 
$$\nabla_\alpha\sigma=2\mathbf T^\beta\mathcal F_{\beta\alpha}.$$
$\sigma$ is normalized so that $\sigma=f+\mathrm{i}\psi\to1$ as $|x|\to\infty$, where $f=-\mathbf g(\mathbf T,\mathbf T)$. We write
$\mathcal F^2=\mathcal F_{\alpha\beta}\mathcal F^{\alpha\beta}$ for the
complex bilinear contraction.

The main theorem of this paper is
\begin{theorem}\label{thm:main}
Suppose that $(\mathbf{M},\mathbf{g})$ is a regular non-degenerate stationary vacuum spacetime with
$\mathbf T|_S\not\equiv0$. Assume moreover that the portion of $\Sigma_0$ for $|x|\ge1$ is maximal.  Then
\begin{equation}\label{eq:technicalassumption}
 -4M^2\mathcal F^2=(1-\sigma)^4\quad\text{on }S,\qquad
 \mathrm{Re}\frac{1}{1-\sigma}>\frac12\quad\text{at some point}\ p\in S.
\end{equation}
Consequently, by \cite{IonescuKlainerman2009}, $\mathbf{E}$ is isometric to the exterior of some Kerr spacetime with mass $M$ and angular momentum $J=Ma$ with $0<|J|<M^2$. 
\end{theorem}
\begin{remark}
The existence of a maximal slice in a regular non-degenerate stationary vacuum spacetime can be obtained by \cite{ChruscielWald1994}, and the required decay in the above assumptions may be further obtained by elliptic estimates. 
\end{remark}
\begin{remark}
If $\mathbf{T}|_S\equiv0$, by constructing a maximal slice, the spacetime can be shown to be static and hence $\mathbf{E}$ is isometric to the exterior of Schwarzschild, see \cite{ChruscielCosta2008}.
\end{remark}
\begin{remark}
The maximality of $\Sigma_0$ in our assumption implies that $\mathbf{E}=\mathbb{R}\times \Sigma_0$. In fact, write $\mathbf T=Nn+X$ on $\Sigma_0$, where $n$ is the future unit normal,
$X$ is tangent, $h=\mathbf g|_{T\Sigma_0}$ and
$k(U,V)=-\mathbf g(\nabla_U n,V)$. The lapse equation is
$$
 \Delta_hN=|k|_h^2N,\qquad N|_S=0,\qquad N\to1.
$$
The maximum principle gives $0<N<1$ in the interior. This means that $-\mathbf{g}(n,N)>0$ and is needed in applying the uniqueness theorem in \cite{IonescuKlainerman2009}.  This also means that the map $\mathbb{R}\times\Sigma_0\to \mathbf E$ by $(t,x)\to\varphi^{\mathbf{T}}_t(x)$, the flow of $\mathbf{T}$, is local diffeomorphism. Surjectivity is a consequence of the completeness assumption, and injectivity follows by global hyperbolicity.

Finally we have
\begin{equation}\label{eq:f-less-one}
 f:=-\mathbf g(\mathbf T,\mathbf T)=N^2-|X|_h^2<1.
\end{equation}

\end{remark}

\subsection{Degenerate horizons and multiple black holes}

A degenerate event horizon has $\kappa=0$ and no regular bifurcation
sphere for its generator. If a connected degenerate horizon is rotating,
analytic rigidity gives a global axial Killing field under the usual
$I^+$-regularity hypotheses in four dimensions
\cite{ChruscielCosta2008}. In vacuum the axisymmetric uniqueness
theorems then identify the exterior with extremal Kerr
\cite{AmselHorowitzMarolfRoberts2010,FiguerasLucietti2010}. Chru\'sciel--Nguyen proved the corresponding 
Kerr--Newman uniqueness theorem for a connected rotating degenerate horizon in an
$I^+$-regular analytic electrovacuum spacetime \cite{ChruscielNguyen2010}. If the single
vacuum horizon is non-rotating, the Komar identity gives $M=\kappa A/(4\pi)$,
where $A$ is the area of a horizon cross-section. Degeneracy would give
$M=0$, which the rigidity case of the positive mass theorem rules out for
a regular black hole \cite{ChruscielCosta2008}. Together with the non-degenerate
results discussed above, one can conclude that a stationary, asymptotically flat, $I^+$-regular, analytic vacuum spacetime with a connected horizon must be the Kerr spacetime. It is still interesting to study uniqueness of degenerate rotating black hole without analyticity assumption. 

A physically relevant case is that the spacetime has multiple black holes, i.e., the horizon has several components. 
Intuitively, since gravitational interaction gives rise to an attractive effect between black holes, the existence of multiple black holes would violate the stationarity of the spacetime. Thus it is also a long time conjecture that multiple black holes cannot occur in stationary spacetimes.
If at least one component
is rotating, analytic rigidity again gives global axisymmetry
\cite{ChruscielCosta2008}. For a stationary and axisymmetric spacetime, by an agular momentum-area relation proved by Hennig--Ansorg--Cederbaum  \cite{HennigCederbaumAnsorg2010}, and by the recent work 
of Han--Khuri--Weinstein--Xiong, one can show that every
finite axis interval between black holes in such a stationary axisymmetric
vacuum configuration has a conical singularity \cite{HanKhuriWeinsteinXiong2026},
contradicting regularity of the axis. Also see an earlier work on the impossibility of two black holes \cite{NeugebauerHennig2012}.
If all components are non-rotating and non-degenerate, the maximal-slice
staticity argument still holds \cite{ChruscielCosta2008}, and the classical static black hole uniqueness theorem by Bunting--Masood--ul--Alam~\cite{BuntingMasoodulAlam1987} tells that the spacetime must be Schwarzschild. The final possible case is that all components are non-rotating but at least one of them is degenerate, which is excluded by the recent work of
Huang--Li--Zhu \cite{HuangLiZhu2026}. 
Consequently, one can conclude that multiple black holes cannot occur in a stationary, asymptotically flat, $I^+$-regular, analytic vacuum spacetime.

It would be interesting to consider whether multiple black holes exist or not in a stationary, asymptotically flat, $I^+$-regular, \emph{smooth} vacuum spacetime. We mention the work of Wong--Yu \cite{WongYu2014} which excludes the possibility of smooth stationary multiple black holes if certain smallness is imposed on a Cauchy hypersurface in vacuum and electrovacuum cases. The connectedness assumption on the horizon in \cite{AlexakisIonescuKlainerman2010b} is therefore not needed. 

\subsection{The main idea}

The proof is based on the framework established in \cite{Radu2026b}. By using the Harrison transformation \cite{Harrison1968}, one can defined a family of stationary electrovacuum spacetime
$$
 \mathbf g_q=\frac14\left(
 D_q\mathbf g+\mathbf T^\flat\otimes W_q+W_q\otimes\mathbf T^\flat
 -\frac f{D_q}W_q\otimes W_q\right)
$$
for $q\in[0,1]$, where $D_q=|q(1+\sigma)+(1-\sigma)|^2$ and $W_q$ is a suitably selected one-form with its spatial components decay like $(q+r^{-1})O(r^{-3})$. The complex electromagnetic two-form is
\begin{equation*}
 \mathcal{H}_q
=-\sqrt{1-q^2}
 \frac{q(1+\overline\sigma)+(1-\overline\sigma)}
      {q(1+\sigma)+(1-\sigma)}
 \left(\mathcal{F}+\frac{W_q\wedge\D\sigma}{2D_q}\right)
 \end{equation*}
 and the Maxwell two-forms 
$$
 \mathbf F_q=\operatorname{Re}\mathcal{H}_q, \quad {}^{\star_{\mathbf g_q}}\mathbf F_q=\operatorname{Im}\mathcal{H}_q.$$
 The bifurcation horizon structure is maintained under this transformation.  By the decay of $\sigma$, $D_q=4(q+r^{-1})^2(1+O_5(r^{-1}))$ with the implicit constants independent of $q\in[0,1]$. All $\mathbf{g}_q$ for $q>0$ are asymptotically flat in a normalized coordinates system $t_q\asymp t/q $ and $x_q=qx$, but the asymptotically flat ends recede to infinity in the ambient spacetime manifold.  The metric $\mathbf{g}_0$ is then no longer asymptotically flat and its spatial sections are asymptotically cylindrical. This can also be seen from the expression of $D_q$ if we let $q\to0$ with $r$ fixed.  Under the mass-one normalization of the original ADM mass, the asymptotic area of the spatial sections of $\mathbf{g}_0$ are $4\pi$.

The strategy of Radu \cite{Radu2026b} is then show that a sphere $C$ in the original asymptotic region has area lower bound $4\pi$ measured in $\mathbf{g}_0$, by comparing to a stable minimal $2$-sphere in a maximal slice bounded by the bifurcate sphere $S$ and $C$. However, since $\mathbf{g}_0$ is asymptotically cylindrical, constructing a maximal slice in $\mathbf{g}_0$ may fail. So the comparison is done for a family of $C_q$ in $\mathbf{g}_q$ for all sufficiently small $q$, and letting $q\downarrow0$.  The area of this stable minimal $2$-sphere is bounded below by $4\pi(1-q^2)$, which is exactly the square of the integral of ${}^{\star_{\mathbf g_q}}\mathbf F_q$ over a sphere homologous to $S$ divided by $4\pi$.  In his construction, $\mathbf{T}|_{C_q}$ should be timelike, which is used to control the slope of the maximal slice at $C_q$.  By carefully arrange the locations of these spheres, it can be shown that the $\mathbf{g}_0$-future null cone of the limit sphere $C=\lim_{q\downarrow 0}C_q$ extends to infinity without focal points and cut points, with limit area $4\pi$. Therefore, by Raychaudhuri equation in $\mathbf{g}_0$,  the area of $C$ in $\mathbf{g}_0$ is exactly $4\pi$ and this null cone is a non-expanding horizon and because of the dominant energy condition for $\mathbf{g}_0$. The null cone then has vanishing shear. By stationarity Radu identifies the transformed asymptotic region is the Bertotti–Robinson solution whose inverse transformation is exactly Kerr.  No smallness of the angular momentum is imposed in this step. By imposing smallness, Radu was able to show that a smallness condition holds on the bifurcate sphere and then \cite{AlexakisIonescuKlainerman2014} applies. 

Our strategy must overcome the difficulties caused by the failure of $\mathbf T$ to be timelike in the ergoregion. We choose to bound the area of $C$ from below  directly by the area of the bifurcate sphere $S$ instead of the stable minimal surface.  The idea is direct: by translating $C$ along the flow of the stationarity  $\mathbf{T}$ for sufficiently large time, its inner boundary of causal past is arbitrary close to $S$.   By Raychaudhuri again, the area of $C$ is bounded below by the area of $S$, up to an error tending to zero when $t\to\infty$.  However, the above heuristic arguments require that $\mathbf{g}_0$ has a good causal structure, which is not known a priori.  It can only be shown that all $\mathbf{g}_q$ for $q>0$ are globally hyperbolic, by constructing asymptotically flat maximal slices with boundary $S$. The construction on compact regions is similar to that in \cite{Radu2026b}, while additional work is needed to obtain estimates at infinity.  Then we carefully choose $C_q$ with area $4\pi+o_q(1)$ and prove that the area of $S$ is again bounded from below by $4\pi(1-q^2)$, utilizing the fact that $S$ is a stable MOTS in $\mathbf{g}_q$ and the area-charge inequality. In our proof, $C_q$ is not required to have a limit position. We have therefore obtained
$$
4\pi(1-q^2)\le |S|_{\mathbf g_q}
\le |C_q|_{\mathbf g_q}=4\pi+o_q(1).
$$
 By letting $q\downarrow0$, we are able to show that the areas of $S$ in $\mathbf{g}_0$ is $4\pi$. The equality of the area-charge inequality also shows that the electric density of $\mathbf{F}_0$ is constant $1$ and magnetic density is constant $0$. Remarkably, these identities  imply precisely the   technical assumptions \eqref{eq:technicalassumption} of Ionescu--Klainerman~\cite{IonescuKlainerman2009}, allowing us to invoke their rigidity theorem to identify the Kerr exterior.  We remark that the stability of the minimal surface in \cite{Radu2026b} and of the bifurcate sphere $S$ in our proof are crucial in the proof to obtain area lower bound.  This use of stability together with the Gauss equation and the Gauss--Bonnet theorem goes back to the work of Schoen--Yau \cite{SchoenYau1979a,SchoenYau1979b}.  We also note that the underlying rigidity mechanism is reminiscent of that in Israel’s work \cite{Israel1967} on static black holes.

\subsection*{Acknowledgement}

 The authors are supported by National Key R\&D Program of China (No. 2022YFA1005400, No. 2024YFA1015000) and NSFC (12326602, 12141106).
 
 \subsection*{AI Disclosure}
AI tools are used to assist in the development of the paper, including  checking calculations,  editing and polishing the text, and the clarification and verification of certain technical arguments in the construction of maximal slices. 
\section{The Harrison transformation}\label{sec:harrison}

We will normalize the mass so that $M=1$ throughout the paper.

\subsection{The regular formulas}

Let $\mathbf T$ be the stationary Killing field of the vacuum metric
$\mathbf g$, and write $\mathbf E$ for the open domain of outer
communications, $S$ for the bifurcation sphere, and $\Sigma_1$ for the
original maximal section. We use signature $(-,+,+,+)$ and normalize the
mass to one and $\mathbf g(\mathbf T,\mathbf T)\to-1$ at infinity.
The connection, volume tensor and Hodge star of $\mathbf g$ are denoted
by $\nabla$, $\epsilon$ and ${}^{\star_{\mathbf g}}$. We orient the spacetime so that $dt\wedge dx^1\wedge dx^2\wedge dx^3$ is positive in the asymptotic chart, and use this orientation
for all Hodge stars.

Define the Killing two-form and its complex form by
$$
 F_{\alpha\beta}=\nabla_\alpha\mathbf T_\beta,\quad
 \mathcal{F}_{\alpha\beta}
 =F_{\alpha\beta}+\frac{\mathrm i}{2}
       \epsilon_{\alpha\beta\mu\nu}F^{\mu\nu}.
$$
Indices here are raised and lowered using $\mathbf g$, and
${}^{\star_{\mathbf g}}\mathcal{F}=-\mathrm i\mathcal{F}$.
The normalized vacuum Ernst potential is
$$
 \nabla_\alpha\sigma=2\mathbf T^\beta\mathcal{F}_{\beta\alpha},
 \quad \sigma=f+\mathrm i\psi\to1.
$$
In particular,
$\nabla_\alpha\psi=\mathbf T^\beta\epsilon_{\beta\alpha\mu\nu}F^{\mu\nu}$.
The defining one-form is closed by the vacuum equations. 

For $q\in[0,1]$, put
\begin{equation}\label{eq:Dq-definition}
 D_q=|q(1+\sigma)+(1-\sigma)|^2.
\end{equation}
The inequality \eqref{eq:f-less-one} gives
$$
 \operatorname{Re}\bigl(q(1+\sigma)+(1-\sigma)\bigr)
 =2q+(1-q)(1-f)>0\quad\hbox{on }\mathbf E.
$$

We will choose a stationary one-form $W_q$ below, verifying the equation
\begin{equation}\label{eq:wq-identities}
 \begin{split}
 \D W_q&=4(1-q)\bigl(((1-q)f-(1+q))F
                  +(1-q)\psi\,{}^{\star_{\mathbf g}}F\bigr),\\
 W_q(\mathbf T)&=d_q-D_q.
 \end{split}
\end{equation}
Here $d_q=D_q(p)$ where $p\in S$ is the zero of $\mathbf{T}$ on $S$, the pole of the local
circle action. The normalization in the second line is forced by smoothness at $S$. To see this, Cartan's identity gives
$$
 \D\bigl(W_q(\mathbf T)+D_q\bigr)
 =\mathcal{L}_{\mathbf T}W_q-i_{\mathbf T}\D W_q+\D D_q=0.
$$
Then $W_q(\mathbf T)+D_q$ is constant on $S$. At the zero $p$ of $\mathbf{T}$, $W_q(\mathbf{T})$ must be zero.

The regular family Harrison metric for $q\in[0,1]$ is
\begin{equation}\label{eq:regular-metric}
 \mathbf g_q=\frac14\left(
 D_q\mathbf g+\mathbf T^\flat\otimes W_q+W_q\otimes\mathbf T^\flat
 -\frac f{D_q}W_q\otimes W_q\right).
\end{equation}
Equivalently, define the linear map on each tangent space by
\begin{equation}\label{eq:harrison-metric}
 A_qv=v+\frac{W_q(v)}{D_q}\mathbf T,\quad
 \mathbf g_q=\frac{D_q}{4}A_q^*\mathbf g.
\end{equation}
The contraction in \eqref{eq:wq-identities} also gives
\begin{equation}\label{eq:aq-inverse}
 \det A_q=\frac{d_q}{D_q}>0,\quad
 A_q^{-1}(w)=w-\frac{W_q(w)}{d_q}\mathbf T.
\end{equation}
Consequently the regular metrics are Lorentzian, with
$$
 \mathbf g_1=\mathbf g,\quad
 \mathcal{L}_{\mathbf T}\mathbf g_q=0,\quad
 \mathbf g_q(\mathbf T,\mathbf T)
 =\frac{d_q^2}{4D_q}\mathbf g(\mathbf T,\mathbf T).
$$

The complex electromagnetic two-form is
\begin{equation}\label{eq:transformed-two-form}
 \begin{split}
 \mathcal{H}_q
 &=-\sqrt{1-q^2}
 \frac{q(1+\overline\sigma)+(1-\overline\sigma)}
      {q(1+\sigma)+(1-\sigma)}A_q^*\mathcal{F}\\
 &=-\sqrt{1-q^2}
 \frac{q(1+\overline\sigma)+(1-\overline\sigma)}
      {q(1+\sigma)+(1-\sigma)}
 \left(\mathcal{F}+\frac{W_q\wedge\D\sigma}{2D_q}\right)
 \end{split}
 \end{equation}
 and the Maxwell two-forms 
$$
 \mathbf F_q=\operatorname{Re}\mathcal{H}_q, \quad {}^{\star_{\mathbf g_q}}\mathbf F_q=\operatorname{Im}\mathcal{H}_q.$$
These formulas have no division by $f$. Since $\mathbf T$ is tangent to
$S$, $f|_S\le0$ and $D_q|_S\ge(1+q)^2$; they therefore also apply in a
regular neighborhood of $S$ once $W_q$ is chosen smooth there.

\begin{proposition}[{\cite[Proposition 3.1]{Radu2026b}}]\label{prop:harrison-family}
The fields $(\mathbf g_q,\mathbf F_q)$ satisfy
\begin{equation}\label{eq:einstein-maxwell}
 \begin{split}
 \operatorname{Ric}(\mathbf g_q)_{\alpha\beta}
 &=2\left((\mathbf F_q)_{\alpha\mu}(\mathbf F_q)_\beta{}^\mu
 -\frac14(\mathbf g_q)_{\alpha\beta}
              (\mathbf F_q)_{\mu\nu}(\mathbf F_q)^{\mu\nu}\right),\\
 \D\mathbf F_q&=0,\quad
 \D\bigl({}^{\star_{\mathbf g_q}}\mathbf F_q\bigr)=0,
 \end{split}
\end{equation}
on $\mathbf E$ and in their extension near $S$. Indices in
\eqref{eq:einstein-maxwell} are raised using $\mathbf g_q$.
The metrics are jointly smooth in spacetime and $q\in[0,1]$; the
two-forms are smooth in spacetime and continuous in $q$, with
$(\mathbf g_1,\mathbf F_1)=(\mathbf g,0)$. With the sphere orientation
inherited from the asymptotic end,
\begin{equation}\label{eq:charge}
 \frac1{4\pi}\int_S{}^{\star_{\mathbf g_q}}\mathbf F_q=\sqrt{1-q^2},\quad
 \frac1{4\pi}\int_S\mathbf F_q=0.
\end{equation}
\end{proposition}

\subsection{The choice of $W_q$}\label{sec:harrison-primitive}

The whole family of $W_q$ can be determined by a single choice of $W_0$:
\begin{equation}\label{eq:wq}
 W_q=(1-q)^2W_0-4q(1-q)\mathbf T^\flat.
\end{equation}
and 
$$
 \Theta_0=\D W_0=4\bigl((f-1)F+\psi\,{}^{\star_{\mathbf g}}F\bigr),
$$

$W_0$ is chosen regular near $S$ and with decaying spatial components
at infinity. In the original stationary coordinates $(t,x)$, $r=|x|$, 
the mass-one asymptotics are (see \eqref{eq:af})
$$
 \mathbf g_{00}=-1+\frac2r+O_6(r^{-2}),\quad
 \mathbf g_{0i}=O_6(r^{-2}),\quad
 \mathbf g_{ij}=\delta_{ij}+O_6(r^{-1}).
$$
The component definitions give (with $\mathbf{T}=\partial_0$ near infinity)
$$
 f=1-\frac2r+O_6(r^{-2}),\quad
 F_{ij}=\tfrac12(\partial_i\mathbf g_{0j}-\partial_j\mathbf g_{0i})
       =O_5(r^{-3}),\quad
 ({}^{\star_{\mathbf g}}F)_{ij}=O_5(r^{-2}).
$$
Thus $\partial_i\psi=O_5(r^{-3})$, and integration from infinity gives
$$
 \psi=O_6(r^{-2}),
 \quad \sigma=1-\frac2r+O_6(r^{-2}).
$$
It follows that $(\Theta_0)_{ij}=O_5(r^{-4})$. $W_0$ is chosen such that, near infinity
\begin{equation}\label{eq:radial-primitive}
 (W_0)_i(x)=-\int_1^\infty s x^j(\Theta_0)_{ji}(sx)\,\D s
           =O_5(r^{-3}).
\end{equation}
The time component is simply $(W_0)_0=W_0(\mathbf{T})=d_0-D_0$. 
$W_0$ defined above is not necessarily smooth up to $S$.  
For the stationary vacuum metric, the identities $\D F=\D{}^{\star_{\mathbf g}}F=0$, 
and the contraction of 
${}^{\star_{\mathbf g}}F\wedge {}^{\star_{\mathbf g}}F=-F\wedge F$ and $\mathbf{T}$ imply 
$$
 \D\Theta_0=0.
$$
By Stokes theorem, 
$$
\int_{S}\Theta_0
 =\lim_{R\to\infty}\int_{S_R}\Theta_0=0.
$$
Since $H_2(\mathbf E;\mathbb R)$ is generated by $[S_r]$, all periods of $\Theta_0$ vanish. Hence \(\Theta_0=dW_0\) globally. Hence a global  $W_0$ exists.  A suitable choice of $W_0$ near $S$, say $\overline{W}_0$ verify
$$W_0-\overline{W}_0=\D \chi$$
for some function $\chi$ in the overlap region. Then $\overline{W}_0+\D(\eta\chi)$ is the desired one form possessing the asymptotic \eqref{eq:radial-primitive}, where $\eta$ is a stationary cutoff function with value $0$ near $S$ and $1$ near infinity.   See \cite[Appendix A.1]{Radu2026b}.

\subsection{The asymptotic at infinity}\label{sec:harrison-end-asymptotics}

We study the asymptotic of the Harrison family $\mathbf{g}_q$ under the original stationary coordinates $(t,x)$, with $\mathbf T=\partial_t$.
Substitution of
$(W_q)_t=d_q-D_q$ into \eqref{eq:regular-metric} gives
\begin{equation*}
 \begin{split}
 (\mathbf g_q)_{00}&=-\frac{d_q^2f}{4D_q},\\
 (\mathbf g_q)_{0i}&=\frac{d_q}{4}
       \left(\mathbf g_{0i}-\frac f{D_q}(W_q)_i\right),\\
 (\mathbf g_q)_{ij}&=\frac14\left(D_q\mathbf g_{ij}
       +\mathbf g_{0i}(W_q)_j+\mathbf g_{0j}(W_q)_i
       -\frac f{D_q}(W_q)_i(W_q)_j\right).
 \end{split}
\end{equation*}
Using the expansion of $f$, we have
$$
 (1+q)-(1-q)f=2q+\frac{2(1-q)}r+O_5(r^{-2}),
$$
and hence the denominator satisfies
\begin{equation*}
 \begin{split}
 D_q&=[(1+q)-(1-q)f]^2+(1-q)^2\psi^2\\
 &=4\left(q+\frac{1-q}r\right)^2+qO_5(r^{-2})+O_5(r^{-3})\\
 &=4(q+r^{-1})^2
+O_5(q^2r^{-1}+qr^{-2}+r^{-3})
 \end{split}
\end{equation*}
The implicit constants  are independent of
$q\in[0,1]$. Also,
$d_q=D_q(p)=(1+q)^2+(1-q)^2\psi(p)^2$ is bounded above and away
from zero uniformly in $q\in[0,1]$. We will also use a rougher estimate
$$D_q\asymp (q+r^{-1})^2.$$
The time-time component is 
\begin{equation}\label{eq:gq00-asymptotic} (\mathbf g_q)_{00}=-\frac{d_q^2}{16}\left(q+r^{-1}\right)^{-2}+\left(q+r^{-1}\right)^{-4}O_5(q^2r^{-1}+qr^{-2}+r^{-3}).\end{equation}
Also
$$(W_q)_i=(1-q)^2(W_0)_i-4q(1-q)\mathbf{g}_{0i}=(q+r^{-1})O_5(r^{-2}). $$
For the mixed components, we have
\begin{equation}\label{eq:gq0i-asymptotic}
 (\mathbf g_q)_{0i}
 =\frac{d_q}{4D_q}
   \left[D_q\mathbf g_{0i}
                  -f(W_q)_i\right]
 =d_q(q+r^{-1})^{-1}O_5(r^{-2}).
\end{equation}
For the spatial components,  using $D_q\asymp (q+r^{-1})^2$, we have
$$
\mathbf g_{0i}(W_q)_j=O_5(r^{-4}),\quad
\frac f{D_q}(W_q)_i(W_q)_j=O_5(r^{-4}).
$$
So
\begin{equation}\label{eq:gqij-asymptotic}
(\mathbf g_q)_{ij}
=\frac{D_q}{4}\bigl(\delta_{ij}+O_5(r^{-1})\bigr)
+O_5(r^{-4})=\left(q+r^{-1}\right)^2\delta_{ij}
    +O_5(q^2r^{-1}+qr^{-2}+r^{-3}).
\end{equation}

\subsection{Preservation of the bifurcate horizon}\label{sec:horizon}

We next show that the Harrison transformation preserves the bifurcate horizon locally. The argument follows \cite{Radu2026b} and some details will be omitted.

The future and past horizons  $\mathcal{H}^+$ and $\mathcal{H}^-$ are smooth, transverse and nonexpanding near $S$, with $\mathbf{T}$ tangent to them. By the local rigidity theorem of Alexakis--Ionescu--Klainerman \cite{AlexakisIonescuKlainerman2010a}, there is a Hawking Killing field $\mathbf{K}$ in a neighborhood of $S$,
with
$$
 \mathbf{K}=\mathbf{T}+\Omega_H\mathbf{Z},\qquad \mathbf{K}|_S=0,
 \qquad [\mathbf{T},\mathbf{K}]=[\mathbf{T},\mathbf{Z}]=[\mathbf{K},\mathbf{Z}]=0.
$$
where $\Omega_H$ is a constant and  $\mathbf{Z}$ is also Killing and has closed orbits, normalized to have least positive period $2\pi$. It can be shown that there is some constant  $\kappa>0$, called surface gravity  such that on $\mathcal{H}^+$
$$
 \nabla_{\mathbf{K}}\mathbf{K}=\kappa\mathbf{K},\qquad
 \D\bigl(\mathbf{g}(\mathbf{K},\mathbf{K})\bigr)=-2\kappa\mathbf{g}(\mathbf{K},\cdot).
$$
On  $\mathcal{H}^-$  we have $\nabla_{\mathbf K}\mathbf K=-\kappa\mathbf{K}$.  These fields also preserve the the geometry
$$
 \mathcal{L}_{\mathbf{K}}\sigma=\mathcal{L}_{\mathbf{Z}}\sigma=0,
 \qquad
 \mathcal{L}_{\mathbf{K}}\Theta_0=\mathcal{L}_{\mathbf{Z}}\Theta_0=0
$$
Therefore, it is possible to choose a $W_0$ (and hence $W_q$ by \eqref{eq:wq}) which is smooth to $S$ so that
$$\mathcal L_{\mathbf{K}}W_0=0,\quad \mathcal L_{\mathbf{K}}W_q=0.$$
See \cite{Radu2026b} for the precise construction. We will then show the horizons are preserved in $\mathbf{g}_q$ for every $q\in[0,1]$ and $\mathbf{K}$ is still the Hawking Killing field.

By the invariance of $W_q$ in $\mathbf K$, we have, along $\mathcal H^+$, 
$$\mathbf{K}(W_q(\mathbf{K}))=0,$$
and because $\mathbf{K}|_S=0$, we have $W_q(\mathbf{K})=0$ on $\mathcal{H}^+$. Also we have $\mathbf{g}(\mathbf K,\mathbf T)=0$ on $\mathcal{H}^+$ for $\mathbf K$ is tangent to its generators. By \eqref{eq:regular-metric} we have on $\mathcal{H}^+$
$$
 \mathbf{g}_q(\mathbf{K},\cdot)=\frac{D_q}{4}\mathbf{g}(\mathbf{K},\cdot).
$$
It follows that $\mathbf{K}$ is also null normal of $\mathcal{H}^+$ in $\mathbf{g}_q$.  In particular, $\mathcal{H}^+$ is a null hypersurface. Since by construction $\mathcal{L}_{\mathbf{K}}\mathbf{g}_q=0$, $\mathcal{H}^+$ is $\mathbf{g}_q$-Killing horizon.  To check the surface gravity, by \eqref{eq:regular-metric}, we have
$$
 \mathbf{g}_q(\mathbf{K},\mathbf{K})=\frac{D_q}{4}\mathbf{g}(\mathbf{K},\mathbf{K})
 +\frac12 \mathbf{g}(\mathbf{T},\mathbf{K})W_q(\mathbf{K})
 -\frac{f}{4D_q}W_q(\mathbf{K})^2.
$$
in a neighborhood of $S$. Taking differential and restricted on $\mathcal H^+$ we have
$$\D\left( \mathbf{g}_q(\mathbf{K},\mathbf{K})\right)=\frac{D_q}{4}\D\left(\mathbf{g}(\mathbf{K},\mathbf{K})\right)=-\frac{D_q}{4}\cdot 2\kappa\mathbf{g}(\mathbf{K},\cdot)=-2\kappa\mathbf{g}_q(\mathbf{K},\cdot),$$
which shows that $\kappa$ is the $\mathbf{g}_q$-surface gravity of $\mathcal H^+$. Similar arguments work for $\mathcal H^-$. Because both $\mathcal{H}^{\pm}$ are Killing, and then the null second fundamental forms of $S$ vanish,  $S$ is $\mathbf{g}_q$-totally geodeisc.

\section{Complete maximal hypersurfaces for $q>0$}\label{app:complete-maximal}

In this section we construct maximal slices for $\mathbf{g}_q$ for $q>0$.   They are constructed as a graph on the original maximal slice, complete up to their physical boundary $S$. All estimates below are uniform for $q\in[\delta,1]$, with $\delta>0$ fixed. The construction basically follows the line of that in \cite{Radu2026b}, with some modifications needed to handle the asymptotic end.

For $q>0$, the estimates \eqref{eq:gq00-asymptotic}--\eqref{eq:gqij-asymptotic} give
$$
 (\mathbf g_q)_{00}=-\bigl(d_q/(4q)\bigr)^2+O_5(r^{-1}),\quad
 (\mathbf g_q)_{0i}=O_5(r^{-2}),\quad
 (\mathbf g_q)_{ij}=q^2\delta_{ij}+O_5(r^{-1}).
$$
{\bf But note here that the implicit constants depend on $q$. More precisely, depend on $\delta>0$ for all $q\in[\delta,1]$. } Normalize the asymptotically flat coordinates by
$$
 t_q=\frac{d_q}{4q}t,\quad x_q=qx,\quad r_q=qr.
$$
Then
\begin{equation}\label{eq:q_coordinate_decay}
 (\mathbf g_q)_{x_q^0x_q^0}=-1+O_5(r_q^{-1}),\quad
 (\mathbf g_q)_{x_q^0x_q^i}=O_5(r_q^{-2}),\quad
 (\mathbf g_q)_{x_q^ix_q^j}=\delta_{ij}+O_5(r_q^{-1}).
\end{equation}
So $\mathbf g_q$ remains asymptotically flat for every $q>0$. The exterior region $\{|x_q|\ge R\}$ in the normalized coordinates corresponds to $\{|x|\ge R/q\}$, which recedes to infinity in the original coordinates as $q\to0$. But if we fix $\delta>0$, the exterior regions for $q\in[\delta,1]$ is uniform.

We identify
$$\mathbf{E}=\mathbb{R}_t\times\Sigma_1$$
here $t$ is the flow parameter of $\mathbf{T}=\partial_t$ and $\Sigma_1$ represents the original maximal slice $t=0$.  We will then denote the desired maximal slice in $\mathbf{g}_q$ as a graph over $\Sigma_1$ along the $\mathbf T$-flow
$$
 \Sigma_q^\circ=\{(u_q(x),x):x\in\Sigma_1^\circ\}.
$$
The boundary condition will be specified later. On the end, we use the normalized coordinates $(t_q,x_q)$ and the asymptotics \eqref{eq:q_coordinate_decay} above. We define $\widetilde{u}_q$ by writing the end of $\Sigma_q$ as the stationary graph $t_q=\widetilde{u}_q(x_q)$, that is
$$ \widetilde{u}_q(qx)=\frac{d_q}{4q}u_q(x).$$
We require $\widetilde{u}_q\to0$ in this normalization. For any fixed $\delta>0$, the estimates for $\widetilde{u}_q$ is equivalent to $u_q$ for $q\in[\delta,1]$ since $qx=x_q\le x$. Dependence on $\delta$ in the estimates are allowed. In this section we prove the following. 

\begin{theorem}\label{thm:complete-maximal}
For every $\delta\in(0,1)$ and $q\in[\delta,1]$, there exists a smooth  $\mathbf{g}_q$-maximal slice $\Sigma_q$ with boundary $S$, and its interior $\Sigma_q^\circ$ is a graph over $\Sigma_1^\circ$ along the $\mathbf T$-flow, with the following estimates at the asymptotic end flat
\begin{equation}\label{eq:complete-maximal-end-decay}
 |\partial_{x_q}^j\widetilde{u}_q|\le C_{\delta,j}r_q^{-3/4-j},\qquad0\le j\le4.
\end{equation}
\end{theorem}

 Clearly $q=1$ we have a maximal slice which is the original one. The proof is to show that, for every $\delta\in(0,1)$ the set of numbers $q\in [\delta,1]$ verifying the conclusion of the above theorem is open and closed in $[\delta,1]$. 

\subsection{Openness}

We define the weighted H\"older spaces using the original asymptotically flat coordinates $x$ on
$\Sigma_1$, with $r=|x|$. Fix a compact set $K=\Sigma_1\backslash \{r>2r_0\}$, and set
$A_R=\{R<r<2R\}$ for $R\ge r_0$. For $k\ge0$, $0<\alpha<1$ and $\beta\in\mathbb{R}$, we use the norm
$$
 \|w\|_{C^{k,\alpha}_\beta}
 =\|w\|_{C^{k,\alpha}(K)}
 +\sup_{R\ge r_0}R^{-\beta}
 \left(\sum_{j=0}^k R^j\sup_{A_R}|\partial_x^jw|
       +R^{k+\alpha}[\partial^kw]_{\alpha;A_R}\right),
$$
where 
$[v]_{\alpha;A_R}=\sup_{x\ne y\in A_R}\frac{|v(x)-v(y)|}{|x-y|^\alpha}$. 

\begin{lemma}
For a smooth asymptotically flat exterior metric $h-\delta\in C^{3,\alpha}_{-1}$ with compact boundary $S$ and $V\ge0$, $V\in C^{2,\alpha}_{-4}$,
$$
 L=\Delta_h-V:\{w\in C^{4,\alpha}_\beta:w|_S=0\}
 \to C^{2,\alpha}_{\beta-2}
$$
is an isomorphism whenever $-1<\beta<0$.
\end{lemma}

\begin{proof}
The argument follows basically the same to that in \cite[Lemma 3.3]{AnHuang2024} by Fredholm theory. See also the case without inner boudnary in \cite[Sections~3--4]{ChaljubSimonChoquetBruhat1979}. We use the fact that the Fredholm index of $\Delta_h$ is $0$ when $-1=2-3<\beta<0$ and $w\mapsto Vw$ is compact as an operator $C^{4,\alpha}_\beta\to C^{2,\alpha}_{\beta-2}$. The nonnegativity of $V$ shows that $L$ has empty kernel by maximum principle. 
\end{proof}

Now suppose that $q_*>\delta$ verifies the conclusion of Theorem \ref{thm:complete-maximal}. 
Let $\Sigma_{q_*}$ be the maximal slice with induce metric $h_{q_*}$ in $\mathbf{g}_{q_*}$  and choose a future timelike transverse field $Z$, equal to its unit normal $n_{q_*}$ on the compact part and to $\mathbf T$ on the asymptotic flat end. The continued maximal slice is constructed as a graph $w$ along $Z$-flow  over $\Sigma_{q_*}$.   In this flow chart we impose $w|_S=0$.  Denote $\mathscr{H}(q, w)$ be the mean curvature of the graph of $w$ in this flow chart in $\mathbf{g}_q$, its linearization with respect to $w$ reads (see \cite[P. 170]{Bartnik1988})
$$L_*(a_*w)=(\Delta_{h_{q_*}}-V_*)(a_*w)=q_*^{-2}(\Delta_{q_*^{-2} h_{q_*}}-q_*^{2}V_*)(a_*w)$$
where  $a_*=-\mathbf g_{q_*}(Z,n_{q_*})>0 $. Note that we cannot simply choose $Z=\mathbf{T}$ since $\mathbf{T}|_S\in TS$ and orthogonal to $n_*$, and
$$ V_*=|k_{q_*}|^2+\operatorname{Ric}(\mathbf g_{q_*})(n_{q_*},n_{q_*})\ge0$$
where ($k_{q_*}$ is the second fundamental form).  Also we have 
$$q_*^{-2}h_*-\delta\in C^{3,\alpha}_{-1}, \quad V_*\in C^{2,\alpha}_{-4}$$
in the fixed coordinates $x$ from the decay \eqref{eq:complete-maximal-end-decay} of $\widetilde{u}_{q_*}$.  We apply the lemma with $\beta=-3/4$. The nonlinear operator is at least $C^1$ into $C^{2,\alpha}_{\beta-2}$. The lemma and implicit function theorem give local continuation.

We will show these nearby slices are in fact graphs of the stationary flow. Their lapse obeys
\begin{equation}\label{eq:lapse-equation}
 \Delta_{h_q}N_q=
 \bigl(|k_{\Sigma_q}|^2+\operatorname{Ric}(\mathbf g_q)(n_q,n_q)\bigr)N_q,\quad
 N_q|_S=0,\quad N_q\to d_q/(4q).
\end{equation}
Strong maximum principle gives
\begin{equation}\label{eq:complete-maximal-lapse-bound}
 0<N_q\le d_q/(4q).
\end{equation}
 So $\mathbf{T}$ is transverse to $\Sigma_q$ since $N_q=-\mathbf{g}_q(\mathbf{T}, n_q)$. Therefore, the projection along the stationary flow
 $$\pi: \Sigma_q^\circ\to \Sigma_{q_*}^\circ$$
has invertible differential and is a local diffeomorphism. We will show that $\pi$ is proper and then it must be a global diffeomorphism, and then the continued maximal slices can be viewed as a $\mathbf{T}$-graph over $\Sigma_0$.  Pick $p_j=\pi(x_j)\to p\in \Sigma_{q_*}^\circ$,  $x_j$ cannot escape to infinity since $Z=\mathbf{T}$ on the asymptotically flat end.  By  applying Hopf lemma to the lapse equation at $q_*$, we have $N_{q_*}\ge c\rho_{q_*}$ where $\rho_{q_*}$ is the distance function to $S$.  On the other hand, $|w|\le \varepsilon\rho_{q_*}$ for small $\varepsilon$ because it is small in $C^1$ and $w|_S=0$.  Therefore, the stationary time $t$ needed for $x_j$ flowing back to $p_j$ is uniformly bounded. Therefore  $x_j$ cannot approach to $S$ and the properness of $\pi$ follows.

\subsection{Closedness}

Now suppose that we have $q_j\downarrow q_*$ with $q_j$ verifying the conclusion of Theorem \ref{thm:complete-maximal}.  We will show that $\Sigma_{q_j}$ is uniformly (in $q_j\in[\delta, 1]$) spacelike and then the maximal slice $\Sigma_{q_*}$ can be obtained by taking suitable limit.

\subsubsection{Estimates near infinity}

 The lapse of the graph $t_q=\widetilde{u}_q(x_q)$ is
$$N_q=\frac{d_q}{4q}\frac{1}{\sqrt{-\mathbf{g}_q^{00}+2\widetilde{u}_i \mathbf{g}_q^{0i}-\widetilde{u}_i\widetilde{u}_j\mathbf{g}_q^{ij}}}$$
where the components of $\mathbf{g}$ is taken in the coordinates  $(t_q,x_q)$. From, \eqref{eq:complete-maximal-lapse-bound} we have
$$-\mathbf{g}_q^{00}+2\widetilde{u}_i \mathbf{g}_q^{0i}-\widetilde{u}_i\widetilde{u}_j\mathbf{g}_q^{ij}\ge 1,$$
by \eqref{eq:q_coordinate_decay} we have
\begin{equation}\label{eq:weak-maximal-end-slope}
  |\partial_{x_q}\widetilde{u}_q|\le C_\delta r_q^{-1/2}.
\end{equation}
This gives the uniform slope bound on the asymptotically flat end.  Thus the sections are uniformly spacelike outside a common radius, and their unit normals are bounded in the asymptotic frame.

Writing $\mathbf{g}=\mathbf{g}_q, h=h_q, \widetilde{u}=\widetilde{u}_q$, the maximal graph equation in $(t_q,x_q)$ is
$$
\partial_i\left(
\frac{\sqrt{-\det \mathbf{g}}\,\bigl(\mathbf{g}^{ij}\widetilde u_j-\mathbf{g}^{i0}\bigr)}
{\sqrt{-\mathbf{g}^{00}+2\mathbf{g}^{0j}\widetilde u_j-\mathbf{g}^{jk}\widetilde u_j \widetilde u_k}}
\right)=0.
$$
Writing in the form
$$
L\widetilde{u}=a^{ij}(\mathbf{g}, \partial\widetilde{u})\widetilde{u}_{ij}+b^j(\mathbf{g},\partial{\mathbf{g}}, \partial \widetilde{u})\widetilde{u}_j=f(\mathbf{g},\partial{\mathbf{g}}),
$$
we will have, in view of the uniform slope estimate \eqref{eq:weak-maximal-end-slope} and the background estimates \eqref{eq:q_coordinate_decay},  
$$a^{ij}=\delta^{ij}+O(r_q^{-1}), \quad
b^j=O(r_q^{-2}),\quad f=O(r_q^{-3}).
$$
In order to apply Schauder estimates, we derive a bound for $\widetilde{u}$. 
For $0<\mu<1$, we have
$$
 L(r_q^{-\mu})\le-c_\mu r_q^{-\mu-2}
$$
beyond a common radius. Set $M_q(r_*)=\max_{r_q=r_*}|\widetilde{u}_q|$ at any sufficiently large fixed $r_*$. Since each $\widetilde{u}_q\to0$, comparison with the positive and negative radial barriers gives
\begin{equation}\label{eq:two-barrier}
 |\widetilde{u}_q(x_q)|\le\left(M_q(r_*)+\frac{K}{r_*}\right)\left(\frac{r_*}{r}\right)^\mu,
 \quad r_q\ge r_*.
\end{equation}
Here $K$ is independent of $q\in [\delta,1]$ and is chosen to dominate the $O(r_q^{-3})$ right-hand side. 

Integrating \eqref{eq:weak-maximal-end-slope} along radial segments from $r_*$ to $4r_*$ gives
$$
 M_q(r_*)\le M_q(4r_*)+2C_\delta\sqrt{r_*}
 \le4^{-\mu}\left(M_q(r_*)+\frac{K}{r_*}\right)+2C_\delta\sqrt{r_*}.
$$
Absorbing $M_q(r_*)$ on the right and  substituting into \eqref{eq:two-barrier} yields
\begin{equation}\label{eq:C0-end}
 |\widetilde{u}_q|\le C_{\delta,\mu}r_q^{-\mu}.
\end{equation}
Here $r_*$ is chosen depending on $\mu$ and $\delta$.  Once we have this $C^0$ bound, by rescaling the the large annuals to a fixed size, by standard  Schauder estimates, we have
$$
 \sum_{j=0}^4 R^j\sup_{R<r_q<2R}|\partial_{x_q}^j\widetilde{u}_q|
 \le C_\delta\left(\sup_{R/2<r_q<4R}|\widetilde{u}_q|+R^{-1}\right).
$$
The term $R^{-1}$ comes from scaling the $f= O(r_q^{-3})$ on the right-hand side of the second-order equation. Taking $\mu=\frac{3}{4}$ proves \eqref{eq:complete-maximal-end-decay}.

\subsubsection{Estimates near the bifurcation sphere}

The remaining estimates concern a neighborhood of $S$ and the middle compact region. We first show a $C^0$ bound for the stationary height $u_q$. 
\begin{lemma} \label{lem:compact-C0}For any compact set $K\subset\Sigma_1^\circ$, we have
$$
 \sup_K|u_q|\le C_{\delta,K}.
$$
\end{lemma}
\begin{proof}Recall from \eqref{eq:harrison-metric}--\eqref{eq:aq-inverse} that
$$
 \mathbf g_q(U,V)=\frac{D_q}{4}\mathbf g(A_qU,A_qV),\quad
 A_q^{-1}V=V-\frac{W_q(V)}{d_q}\mathbf T.
$$
By definition, $A_q$ sends future timelike vectors for $\mathbf g_q$ to future timelike vectors for $\mathbf g$. Choose a smooth future-directed $\mathbf g$-timelike curve $\gamma:[0,1]\to\mathbf E$. Define
$$
 \alpha_q(\lambda)=-\frac1{d_q}\int_0^\lambda
       (W_q)_{\gamma(\mu)}\bigl(\gamma'(\mu)\bigr)\,d\mu,
 \quad
 \gamma_q(\lambda)=\varphi^{\mathbf{T}}_{\alpha_q(\lambda)}
                              \bigl(\gamma(\lambda)\bigr).
$$
Here $\varphi_t^{\mathbf{T}}$ is the flow of $\mathbf{T}$. Thus $ \gamma_q(0)=\gamma(0)$. In the stationary product
coordinates, if $\gamma(\lambda)=(t(\lambda),x(\lambda))$, then
$\gamma_q(\lambda)=(t(\lambda)+\alpha_q(\lambda),x(\lambda))$.
In particular, the two curves follow the same spatial path. We also have
$$
 \gamma_q'(\lambda)
  =(d\varphi^{\mathbf{T}}_{\alpha_q(\lambda)})_{\gamma(\lambda)}
       \bigl(\gamma'(\lambda)+\alpha_q'(\lambda)\mathbf T\bigr)
  =(d\varphi^{\mathbf{T}}_{\alpha_q(\lambda)})_{\gamma(\lambda)}
       \bigl(A_q^{-1}\gamma'(\lambda)\bigr).
$$
 Since the
stationary flow preserves $\mathbf g$, $D_q$ and $A_q$, we obtain
\begin{equation}\label{eq:timelike-gq}
 (\mathbf g_q)_{ \gamma_q(\lambda)}
       ( \gamma_q', \gamma_q')
   =\frac{D_q(\gamma(\lambda))}{4}
       \mathbf g_{\gamma(\lambda)}(\gamma',\gamma')<0.
\end{equation}
Therefore $\gamma_q$ is timelike and future directed.  

Now fix some $p\in K$. pick a $\mathbf{g}_1$-future directed timelike curve  $\gamma:[0,1]\to\mathbf{E}$ from $(t_-, y_-)$ (in the fixed coordinate $(t,x)$) to $(0,p)$ where $\frac{r_*}{q}\le|y_-|\le R_\delta$ for some $R_\delta$ independent of $q$. Then $\gamma_q$ defined above is also a $\mathbf{g}_q$-future directed timelike curve from $(t_-, y_-)$ to $(\alpha_q(1), p)$. 
Along $\gamma_q$, the function $t-u_q(x)$, whose level sets are spacelike and lapse is positive,  is increasing. This is to say
$$t(1)+\alpha_q(1)-u_q(x)> t(0)+\alpha_q(0)-u_q(y_-).$$
We then have
$$u_q(x)<-t_-+\alpha_q(1)+u_q(y_-).$$
By choosing a controlled family of $\gamma$ so that $|t_-|+|\alpha_q(1)|\le C_{\delta,K}$. By \eqref{eq:C0-end} we have $u_q(x)\le C_{\delta,K}$.  The inequality of the other side follows by choosing a future directed timelike curve from $(0,p)$ to asymptotically flat end. 
\end{proof}

We will in the next lemma that the original $\Sigma_1$ is still suitable as a reference near $S$. 

\begin{lemma}\label{lem:reference-spacelike-near-S}
There is a fixed neighborhood $V$ of $S$ in $\Sigma_1$ which is spacelike for every $\mathbf g_q$, $q\in[0,1]$. Write
$$
 \bar h_q=\mathbf g_q|_{T\Sigma_1},
$$
let $\bar n_q$ be the future unit normal of the reference section, and let $\rho$ be the $\bar h_q$-distance to $S$, with outward unit normal $\bar\nu_q=\nabla^{\bar h_q}\rho$ on $S$. The field $\mathbf K$ is future timelike on $V\setminus S$, and
$$
 \mathbf K|_S=0,\quad
 \nabla^{(q)}_{\bar\nu_q}\mathbf K=\kappa\bar n_q
 \quad\text{on }S.
$$
Moreover, the map $(s,x)\mapsto\varphi_s^{\mathbf K}(x)$ is a diffeomorphism from $\mathbb R\times(V\setminus S)$ to its image, where $\varphi_s^{\mathbf K}$ is the flow of $\mathbf K$.
\end{lemma}

\begin{proof}
We first extend $\Sigma_1$ smoothly through $S$; the extension need not be maximal. At $S$, $\mathbf T\in TS$, so the invertible map $A_q$ in \eqref{eq:harrison-metric} preserves $T\Sigma_1$. Hence $\mathbf g_q|_{T\Sigma_1}$ is positive definite there. Smooth dependence and compactness of $[0,1]\times S$ give a common spacelike neighborhood in the extended section. Its induced metrics, unit normals and normal exponential maps have uniform smooth bounds.

Since $\mathbf K$ vanishes on $S$, $\nabla_Y^{(q)}\mathbf K=0$ for $Y\in TS$. The Killing equation therefore makes $\nabla^{(q)}_{\bar\nu_q}\mathbf K$ parallel to $\bar n_q$, with coefficient $\kappa$ given by the fixed normal boost. Taylor expansion along the outward normal geodesics in the reference section gives
$$
 \mathbf K=\kappa\rho\,\bar n_q+O(\rho^2),\quad
 \mathbf g_q(\mathbf K,\mathbf K)
 =-\kappa^2\rho^2+O(\rho^3).
$$
The remainders are uniform for $q\in[0,1]$. Since $\kappa>0$, $\mathbf K$ is future timelike for small $\rho>0$. In particular, $\mathbf{K}$ is transverse to $V\backslash S$, hence $\varphi^{\mathbf{K}}_s$ is a local diffeomorphism.  The injectivity follows from the fact that $\mathbf{K}$ is timelike and $\Sigma_1$ is a Cauchy hypersurface (in $\mathbf{g}_1$).
\end{proof}

\begin{lemma}\label{lem:maximal-boundary-estimates}
Fix $\delta>0$. In the neighborhood $V$ of Lemma \ref{lem:reference-spacelike-near-S}, the maximal sections in with  $q\in[\delta,1]$ are graphs along the $\mathbf K$ flow over $\Sigma_1$, with uniformly bounded flow heights near $S$. By shrinking $V$, they are uniformly spacelike graphs along the normal geodesic flow over $\Sigma_1$, with zero boundary values on $S$ and uniform $C^{k,\alpha}$ bounds. \end{lemma}

\begin{proof}
For $x\in V\backslash S$, define the flow height $u_q^{\mathbf{K}}(x)$ by
$$
 \varphi_{u_q^{\mathbf{K}}(x)}^{\mathbf K}x\in\Sigma_q.
$$
$u_q^{\mathbf{K}}$ is well-defined because, $t-u_q(x)$ strictly increases along the future timelike $\mathbf K$ orbit and
$$
 (t-u_q)(\varphi_s^{\mathbf K}x)
 =s+(t-u_q)(\varphi_{\Omega_Hs}^{\mathbf Z}x)
$$
by the flow composition $\varphi_s^{\mathbf K}=\varphi_s^T\circ\varphi_{\Omega_H s}^{\mathbf Z}$ since  $\mathbf{K}=\mathbf{T}+\Omega_H\mathbf{Z}, [\mathbf{T},\mathbf{K}]=0$. The values of $(t-u_q)(\varphi_{\Omega_Hs}^{\mathbf Z}x)$ is uniformly bounded on a circle, then by intermediate value theorem, the must be (exactly one) $s\in\mathbb{R}$ such that $ (t-u_q)(\varphi_s^{\mathbf K}x)=0$. Moreover, fix a compact set $K\Subset \Sigma_1^\circ$,  the set $\varphi^{\mathbf{T}}_{-u_q(x)}\circ\varphi^{\mathbf{K}}_{u^{\mathbf K}(x)}(K)$ is still compact in $\Sigma_1^\circ$. By Lemma \ref{lem:compact-C0}, $|u_q^{\mathbf{K}}|\le C_{\delta,K}$ on $K$.

To extend $u_q^{\mathbf{K}}$ smoothly to $S$, we consider $\Sigma_1$ and $\Sigma_q$ as image of the map 
\begin{equation*}
\begin{split}\Sigma_1: (\varrho,\vartheta)\mapsto \exp^{(q)}_{\vartheta}\left(\varrho(\cosh\alpha_{q}\bar{\nu}_q+\sinh\alpha_{q}\bar{n}_q)\right)\\
\Sigma_q: (\varrho,\vartheta)\mapsto \exp^{(q)}_{\vartheta}\left(\varrho(\cosh\beta_q\bar{\nu}_q+\sinh\beta_q\bar{n}_q)\right)
\end{split}
\end{equation*}
where $\varrho>0$ small and $\vartheta\in S$. $\alpha_q=\alpha_q(\varrho,\vartheta),\beta_q=(\varrho,\vartheta)$ are smooth functions. $\mathbf{K}$ is Killing, it commute with the exponential map. By Lemma \ref{lem:reference-spacelike-near-S}, we have, for all $s$,
 \begin{equation}\label{eq:boost-angle}
\D\varphi_s^{\mathbf K}
\bigl(\cosh\alpha_q\,\bar\nu_q+\sinh\alpha_q\,\bar n_q\bigr)
=
\cosh(\alpha_q+\kappa s)\bar\nu_q
+\sinh(\alpha_q+\kappa s)\bar n_q.
\end{equation}
We then have
$$
\alpha_q +\kappa u_q^{\mathbf K}
=\beta_q.
$$
So $u_q^{\mathbf{K}}\big|_S=\frac{\beta_q-\alpha_q}{\kappa}=\frac{\beta_q(0,\vartheta)}{\kappa}$ since $\alpha_q\big|_S=0$.    This $\beta_q$ may tend to infinity as $q$ varies, below we will show that $\beta_q$ is uniformly bounded for $q\in[\delta,1]$. 

We construct barriers to obtain $C^0$ estimate.  Suppressing $q$, and fix $\rho_0>0$ small enough, put
$$
 B_c^\pm=
 \left\{\varphi^{\mathbf K}_{c\mp M\rho(x)}x:
 x\in\Sigma_1,\ 0\le\rho(x)\le\rho_0\right\}.
$$
All these hypersurfaces have boundary $S$. At $c=0$, their tangent planes and future normals on $S$ agree with those of $\Sigma_1$.  The mean curvature of $B^{\pm}$ is then
$$
 H(B_0^\pm)\big|_S=H(\Sigma_1)\big|_S\mp2\kappa M,
$$
For the barrier comparison we use $H=\operatorname{div}_{\mathbf g_q}n$. Indeed, consider a radial curve $\gamma^{\pm}(\rho)=\varphi^{\mathbf K}_{\mp M\rho}x(\rho)$ on $B^{\pm}_0$ where $x(\rho)$ is a radial curve on $\Sigma_1$ with $x(0)\in S, x'(0)=\bar{\nu}_q$. We have
$$(\gamma^\pm)'(\rho)=\mp M\mathbf{K}(x(\rho))+(\D\varphi^{\mathbf K}_{\mp M\rho})_{x(\rho)}(x'(\rho))$$
and hence $(\gamma^\pm)'(0)=x'(0)=\bar{\nu}_q$. Since $B_0^\pm$ and $\Sigma_1$ share the boundary $S$, their tangent planes agree along $S$. Moreover, from $\mathbf{K}\big|_S=0$, 
$$\nabla^{(q)}_{(\gamma^\pm)'}(\gamma^\pm)'(0)=\nabla^{(q)}_{x'}x'(0)\mp 2M\nabla^{(q)}_{\bar{\nu}_q}\mathbf{K}$$
and the relation of the mean curvatures follows. Choose $M$ independent of $q$ so that 
$$2\kappa M>1+\sup_{q\in[0,1],S}|H(\Sigma_1)|.$$
There exists a small $\rho_0$ on which $B_0^\pm$ are spacelike and $H(B_0^+)<0$, $H(B_0^-)>0$ for all $q\in[0,1]$. Since $B_c^\pm=\varphi_c^{\mathbf K}(B_0^\pm)$ and the flow is isometric, these signs hold for every $c$. Let 
$$
 c_* =\max_{0\le\rho\le\rho_0}(u_q^{\mathbf{K}}+M\rho).
$$
Suppose that $c_*>\max_{\rho=\rho_0}(u_q^{\mathbf{K}}+M\rho_0)$. If the maximum is achieve for some $\rho>0$, then the hypersurface $B_{c_*}^+$ is on the future side of $\Sigma_q$. Moreover, $B_{c_*}^+$ and $\Sigma_q$ are tangent at that point.  By maximum principle, the mean curvature $H(B_0^+)\ge H(\Sigma_q)=0$ at that point, contradicting to the construction of $B_0^+$. If the maximal is achieve for some $p\in S$, $u_q^{\mathbf{K}}(p)=c^*$, the same comparison applies at $p$, since the hypersurfaces are tangent there and the barrier lies on the future side in the exterior half-neighborhood. This shows that
$$\sup_{0\le\rho\le\rho_0}u_q^{\mathbf{K}}(x)+M\rho(x)\le \sup_{\rho=\rho_0}u_q^{\mathbf{K}}(x)+M\rho(x)\le C_{\delta}$$
which is independent of $q\in[\delta,1]$. Lower barrier gives similar bounds and hence we have the $C^0$ estimate
\begin{equation}\label{eq:C0onS}
\sup_{0\le\rho\le\rho_0}|u^{\mathbf{K}}_q|\le C_\delta.
\end{equation}
From \eqref{eq:boost-angle} with $\alpha_q\big|_S=0$, we have
\begin{equation}\label{eq:boost-coordinates}
 \D\varphi_s^{\mathbf K}(\bar\nu_q)
 =\cosh(\kappa s)\bar\nu_q+\sinh(\kappa s)\bar n_q.
\end{equation}
Write $s=u^{\mathbf{K}}_q(x)$, then for $x\in S$, we have 
$$n_q=\D\varphi^{\mathbf K}_{u^{\mathbf{K}}_q(x)}(\bar{n}_q)=\cosh(\kappa u^{\mathbf{K}}_q(x))\bar n_q+\sinh(\kappa u^{\mathbf{K}}_q(x))\bar\nu_q,$$
giving
\begin{equation}\label{eq:boost-bound}
 -\mathbf g_q(n_q,\bar n_q)|_S
 =\cosh(\kappa u_q^{\mathbf{K}}|_S)\le\cosh(\kappa C_\delta).
\end{equation}
Note that $\bar n_q$ has a smooth limit as $q\to0$, so this means that $\Sigma_q$ is uniformaly spacelike over $S$. In the following we will estimate the unit normal outside $S$. 

Near $S$, let $\tau$ be proper time along the future unit normal geodesics from $\Sigma_1$, and let $\pi_\perp$ be the corresponding projection onto $\Sigma_1$. We suppress their dependence on $q$. By \eqref{eq:boost-coordinates},
$$
 \tau(\varphi_s^{\mathbf K}(x))
 =\rho(x)\sinh(\kappa s)+O_s(\rho(x)^2),\qquad
 \rho(\pi_\perp\varphi_s^{\mathbf K}(x))
 =\rho(x)\cosh(\kappa s)+O_s(\rho(x)^2),
$$
with remainders uniform in $q\in[0,1]$ for bounded $s$. Therefore \eqref{eq:C0onS} gives
$$
 |\tau|\le\theta\,\rho\circ\pi_\perp,\quad \theta=\theta_\delta<1
$$
on $\Sigma_q$. This makes $\Sigma_q$ can be viewed as a normal graph over $\Sigma_1$ in the neighborhood of $S$ with its size depending on $\delta$, say $\{0\le\rho\le\rho_\delta\}$. We now estimate the gradient of these graphs. Set
$$
 U=-\operatorname{grad}_{\mathbf g_q}\tau,\quad
 v=-\mathbf g_q(n_q,U)\ge1.
$$
The field $U$ is unit timelike and agrees with $\bar n_q$ on $S$, so \eqref{eq:boost-bound} gives $v\le C_\delta$ on $S$. Differentiating $v$ and using the Codazzi equation and maximality gives a differential inequality on $\Sigma_q$
\begin{equation}\label{eq:compact-maximal-normal-estimate}
 \Delta_{h_q}\log v
 \ge\frac14|\nabla\log v|_{h_q}^2-C_\delta v^2.
\end{equation}
This follows from the similar calculation in \cite[Theorem 3.1]{Bartnik1988} and \cite[Appendix B]{Radu2026b}. 
We choose a cutoff on $\Sigma_q$
$$\chi=\left(1+\frac{\tau-c\rho\circ\pi_\perp}{d}\right)_+$$
with $0<\theta_\delta<c<1$ and $0<d<\frac{1}{2}(c-\theta_\delta)\rho_\delta$. 
$\chi=1$ over $S$,  and  $\chi=0$ on $\rho\circ\pi_\perp=\frac{1}{2}\rho_\delta$.  We compute
$$|\nabla\chi|_{h_q}^2\ge c_\delta v^2-C_\delta,$$
$$|\Delta_{h_q} \chi|\le C_\delta v^2.$$
The latter one use maximality of $\Sigma_q$. 

We apply the maximum principle to $\chi^m v$. At an interior maximum where $\chi>0$,  we have
$$
 \nabla\log v=-m\frac{\nabla\chi}{\chi},
$$
and \eqref{eq:compact-maximal-normal-estimate} gives
$$
 0\ge\Delta_{h_q}\log(\chi^m v)
 \ge\left(\frac{m^2}{4}-m\right)
          \frac{|\nabla\chi|^2}{\chi^2}
       -C_\delta v^2\left(1+\frac m\chi\right).
$$
If $v$ is large enough depending on $\delta$,  $|\nabla\chi|^2\ge\tfrac12c_\delta v^2$. Since $\chi$ is bounded, choosing $m$ sufficiently large makes the right-hand side positive, a contradiction. The cutoff vanishes on the artificial boundary, and $v$ is already bounded on $S$. This establishes the estimates for the unit normal:
$$
 \sup_{\chi\ge \frac{1}{2}} v\le C_\delta.
$$
where $\{\chi\ge\frac{1}{2}\}$ contains a region independent of $q\in[\delta,1]$. 
A finite cover of $S$ gives uniformly spacelike normal graphs on a common neighborhood. Higher order bounds following by Schauder estimates. \end{proof}

\subsubsection{Estimates on compact regions and passage to the limit}

We already have estimates on the asymptotically flat end \eqref{eq:complete-maximal-end-decay} and near a neighborhood (uniform in $q\in[\delta,1]$) from Lemma \ref{lem:maximal-boundary-estimates}. We then consider a compact region $K\Subset \Sigma_1^\circ$, in which we already have $C^0$ bound from Lemma \ref{lem:compact-C0}. We choose $K$ with its inner boundary near $S$ and its outer boundary in the asymptotically flat end, where the unit normals of $\Sigma_q$ are already uniformly controlled. Also we have the global lapse bound \eqref{eq:complete-maximal-lapse-bound}. 

Choose a compact region
$$\widetilde{Q}=\{\varphi_t^T(x):x\in \widetilde{K},\ |t|\le \widetilde{A}\}\Subset\mathbf{E},$$
with $K\Subset \widetilde{K}^\circ$ and $\widetilde{K}$ is compact. Pick a $\mathbf{g}_{q_*}$-timelike curve $\gamma(\lambda)=(t(\lambda), x(\lambda))$ in $\widetilde{Q}$ defined on $\lambda\in[0,1]$. Here $q_*$ is the limit value of $q$.  By compactness, this curve is still timelike for $q_j$ for $j$ sufficiently large. By timelikeness, 
$$\sqrt{-\mathbf{g}_j(\gamma',\gamma')}\le -\mathbf{g}_j(n_j,\gamma')=N_j ((t-u_q)\circ\gamma)')$$
where $n_j=n_{q_j}$ is the future unit normal of $\Sigma_{q_j}$ and is extended to $\widetilde{Q}$ by the $\mathbf T$-flow. The length of $\gamma$ is
$$L_j(\gamma)=\int_0^1\sqrt{-\mathbf{g}_j(\gamma',\gamma')}\D \lambda\le C_\delta (2\widetilde A+C_{\delta,\widetilde K})$$
where the bound from Lemma \ref{lem:compact-C0} is utilized.  Passing to $q_*$, we know that
$$L_*(\gamma)\le C_{\delta,\widetilde A,\widetilde K}$$
for any $\mathbf g_*$-timelike curve lying in $\widetilde{Q}$. Set
$$Q=\{\varphi_t^T(x):x\in K,\ |t|\le A\}\Subset\widetilde{Q}^\circ,$$
where $0<A<\widetilde{A}$. Then, as argued in \cite[Lemma B.2]{Radu2026b}, we can show that any inextendible causal curve lying in a smaller compact region $Q$ can be deformed to timelike curves in $\widetilde{Q}$ with arbitrarily large length, a contradiction.   By Minguzzi \cite[Theorems~4.42 and~4.100]{Minguzzi2019}, there exists a smooth $\mathbf{g}_*$-temporal function $\tau_*$ in (a neighborhood of) $Q$, which is also a $\mathbf{g}_{q_j}$-temporal function for sufficiently large $j$ in $Q$.  By maximum principle, we can show that
$$-\mathbf{g}_j(n_j, -\mathrm{grad}_{\mathbf{g}_j}\tau_*)$$
is uniformly bounded. In fixed local coordinates adapted to $\tau_*$, the sections are uniformly spacelike graphs. The before uniform bound on the angle implies the uniform ellipticity of the maximal graph equations.  The higher order bounds of the graph functions can be obtained by Schauder estimates.  The lapse equation \eqref{eq:lapse-equation}, these uniform local graph estimates and Harnack’s inequality propagate the positive lower bound from the asymptotic end to the compact region:
$$
 N_q\ge c_{\delta,K}>0,\  \text{on}\ \Sigma_q\cap Q.
$$
This lower lapse bound allows the preceding uniform spacelike estimates to be transferred to the stationary graph representation. Then standard Shauder estimates show that 
$$\|u_{q_j}\|_{C^{4,\alpha}(K)}\le C_{\delta,K}.$$

Now let $q_j\to q_*>0$. By the above $C^{4,\alpha}$ bounds, Applying Arzelà–Ascoli on a compact exhaustion of $\Sigma_1^\circ$ gives $u_{q_j}\to u_*\in C_{\mathrm{loc}}^{3,\alpha}$ by passing to a subsequence for some $u_*\in C_{\mathrm{loc}}^{3,\alpha}$.  The limit graph
$$ \Sigma_*^\circ=\{(u_*(x),x):x\in\Sigma_1^\circ\}$$
is spacelike by considering 
$$\mathbf{g}^{-1}_{q_j}(\D(t-u_{q_j}(x)), \D(t-u_{q_j}(x)))=-\frac{1}{N_{q_j}^2}\le -C_{\delta,K}^2.$$
$\Sigma_*^\circ$ is maximal and interior regularity shows that $u^*$ is smooth.  By its construction, $\Sigma_*^\circ$ is graph along the $\mathbf{T}$-flow  over $\Sigma_1$. Near infinity, the asymptotic  \eqref{eq:complete-maximal-end-decay} stills holds for $\widetilde{u}_{q_*}$. Convergence up to $S$ follows from the uniform higher order Hölder estimates in the normal graph representation. This proves Theorem~\ref{thm:complete-maximal}.

\section{Area comparison}\label{sec:minimal-comparison}

For each $q>0$, we are going to identify  a smooth spacelike sphere $C_q$ with positive future outward null expansion, whose stationary projection surrounds $S$, and such that
\begin{equation}\label{eq:sharp-C-area-bound}
 \lim_{q\downarrow0}|C_q|_{\mathbf g_q}=4\pi.
\end{equation}
The spheres may depend on $q$ and recede to infinity if $q\to0$. We choose the spheres to move out to infinity more slowly than the asymptotically flat region.

\begin{proposition}\label{lem:positive-C-expansion}
For a sufficiently large fixed constant $R_0$, set
$$
 C_q=\{t=0,\ r=R_0q^{-2/3}\},\quad 0<q\le1.
$$
in the original coordinates $(t,x)$.  These are smooth embedded spacelike spheres whose stationary projections surround $S$. Their future outward null normals have strictly positive expansion. They satisfy \eqref{eq:sharp-C-area-bound}, their stationary translates form timelike cylinders, and $t$ is temporal on the exterior side of each cylinder.
\end{proposition}

\begin{proof}
 By \eqref{eq:gqij-asymptotic}
$$
(\mathbf g_q)_{ij}
=\frac{D_q}{4}\bigl(\delta_{ij}+O_5(r^{-1})\bigr)
+O_5(r^{-4})=\left(q+r^{-1}\right)^2(\delta_{ij}+O_5(r^{-1})).
$$
So a fixed $R_0$ independent of $q\in[0,1]$ can be chosen so that the induced metric $(\mathbf{g}_q)_{ij}$ on $t=0$ is Riemannian so $t$ is temporal for $r\ge R_0$. The sphere $\{t=0, r=R\ge R_0\}$ is therefore spacelike. The stationary translations form timelike cylinders since
$$ \mathbf g_q(\mathbf T,\mathbf T)
 =\frac{d_q^2}{4D_q}\mathbf g(\mathbf T,\mathbf T)<0$$
 whenever $T$ is timelike in the original metric.  Denote $\gamma_q$ be the induced metric on coordinate sphere. By \eqref{eq:gqij-asymptotic} again, we have
$$
 \gamma_{q,AB}=(1+qr)^2\bigl((\gamma_{\mathbb S^2})_{AB}+O(r^{-1})\bigr),
 \qquad
 \sqrt{\det \gamma_q}=(1+qr)^2\sqrt{\det\gamma_{\mathbb S^2}}
 (1+O(r^{-1})).
$$
Also, we have
$$
 \partial_r\log\sqrt{\det \gamma_q}
 =\frac{2q}{1+qr}+O(r^{-2}).
$$
Choose a future outward null normal $L_q$ and write
$$L_q^\flat=(L_q)_t\D t+(L_q)_r\D r,$$
We normalize $(L_q)_t=\mathbf{g}_q(\mathbf{T}, L_q)=-\frac{d_q}{4}$. The nullity of $L_q$ leads to 
$$\mathbf{g}_q^{tt}\frac{d_q^2}{16}-2\mathbf{g}_q^{tr}\frac{d_q}{4}(L_q)_r+\mathbf{g}_q^{rr}(L_q)_r^2=0,$$
which yields, by \eqref{eq:gq00-asymptotic}, \eqref{eq:gq0i-asymptotic}, \eqref{eq:gqij-asymptotic}
$$
 L_q^r=-\frac{d_q}{4}(\mathbf g_q)^{tr}
+(\mathbf g_q)^{rr}(L_q)_r=\frac{d_q}{4}
 \sqrt{\bigl((\mathbf g_q)^{tr}\bigr)^2
       -(\mathbf g_q)^{tt}(\mathbf g_q)^{rr}}
 =1+O(r^{-1}),
$$
and 
$$(L_q)_r=(q+r^{-1})^2(1+O(r^{-1}))$$
where the positive sign selects the outward normal. 
Again the nullity gives $(L_q)^\alpha (L_q)_\alpha=0$, so $L_q^t (L_q)_t+L_q^r(L_q)_r=0$, and
$$L_q^t=\frac{4}{d_q}L_q^r(L_q)_r=\frac{4}{d_q}(q+r^{-1})^2(1+O(r^{-1})).$$
Finally, 
$$
 L_q^A=(L_q)_t\mathbf{g}_q^{tA}+(L_q)_r\mathbf{g}_q^{rA}=O(r^{-2}).
$$
Using stationarity and $\theta_{L_q}=\frac12\gamma_q^{AB}
(\mathcal{L}_{L_q}\mathbf g_q)_{AB}$, and all estimates for metric coefficients, we obtain
\begin{equation*}
 \begin{split}
 \theta_{L_q}
 =L_q^r\partial_r\log\sqrt{\det \gamma_q}
   +O(r^{-2})=\frac{2q}{1+qr}+O(r^{-2}).
 \end{split}
\end{equation*}
Thus, on $r=R$,
\begin{equation}\label{eq:coordinate-sphere-area-expansion}
 |\{t=0,r=R\}|_{\mathbf g_q}
 =4\pi(1+qR)^2\bigl(1+O(R^{-1})\bigr),\qquad
 \theta_{L_q}=\frac{2q}{1+qR}+O(R^{-2}),
\end{equation}
with implicit constants uniform for $q\in[0,1]$.

We now choose $R=R_0q^{-2/3}$. As $q\to0$,  $R\to\infty$ and
$qR=R_0q^{1/3}\to0$, so the area tends to $4\pi$.
Moreover,
$$
 \frac{2qR^2}{1+qR}
 =\frac{2R_0^2q^{-1/3}}{1+R_0q^{1/3}}
 \ge\frac{2R_0^2}{1+R_0}.
$$
Thus the positive leading term in \eqref{eq:coordinate-sphere-area-expansion}
dominates the error for every $q\in(0,1]$ once $R_0$ is fixed
sufficiently large. 
\end{proof}


For arguments at fixed $q$, we write $C=C_q$, with $C_q$ as in Proposition~\ref{lem:positive-C-expansion}. Its stationary translates form the timelike cylinder
$$
 \mathcal{C}=\bigcup_{t\in\mathbb{R}}\varphi^{\mathbf{T}}_t(C).
$$
We also denote
$$C_t=\varphi^{\mathbf{T}}_t(C).$$

We will prove in this section
\begin{theorem}\label{thm:S<=4pi}
$|S|_{\mathbf{g}_0}\le 4\pi$. 
\end{theorem}

By \eqref{eq:sharp-C-area-bound} and smooth convergence of $\mathbf g_q$ on $S$, it suffices to show $|S|_{\mathbf g_q}\le |C_q|_{\mathbf g_q}$ for sufficiently small $q>0$.  

Fix $q>0$. Theorem~\ref{thm:complete-maximal} gives a complete maximal stationary section $\Sigma_q\subset\mathbf{M}$ with boundary $S$ and interior $\Sigma_q^\circ=\Sigma_q\setminus S\subset\mathbf{E}$. Write $\mathbf{T}=N_qn_q+X_q$ where $X_q\in T\Sigma_q$ and $h_q=\mathbf{g}_q|_{T\Sigma_q}$. Define $v_q:\mathbf{E}\to\mathbb R$ along the integral curve of $\mathbf{T}$, with $\mathbf{T}v_q=1$ and $v_q=0$ on $\Sigma_q^\circ$. Define spatial coordinates $(x^i)$ by translation along $\mathbf{T}$, by ADM $1+3$ decomposition, we have (note \eqref{eq:complete-maximal-lapse-bound})
$$
 \mathbf{g}_q=-N_q^2\D v_q^2+
 h_{q,ij}(\D x^i+X_q^i\D v_q)(\D x^j+X^j_q\D v_q),\quad
 \mathbf{g}_q^{-1}(\D v_q,\D v_q)=-N_q^{-2}<0.
$$
We write $\Sigma_{q,a}$ for the translated hypersurface including its boundary $S$; its interior is $\Sigma_{q,a}^\circ=\{p\in\mathbf{E}:v_q(p)=a\}$. The $v_p$ is temporal and $(\mathbf{E}, \mathbf{g}_q)$ is stably causal for $q\in(0,1]$. We will show that these interiors are Cauchy hypersurfaces and $(\mathbf{E}, \mathbf{g}_q)$ is globally hyperbolic for $q\in(0,1]$.

\begin{lemma}\label{lem:positive-cauchy-foliation}
For every $q\in(0,1]$ and $a\in\mathbb R$, $\Sigma_{q,a}^\circ=\{p\in\mathbf{E}:v_q(p)=a\}$ is a spacelike Cauchy hypersurface of $(\mathbf{E},\mathbf{g}_q)$.
\end{lemma}

\begin{proof}
We suppress $q$ in this proof. Given any inextendible causal curve $\gamma$, parametrized by $(v, x^i(v))$.  It intersects each $\Sigma_v$ at most once. We will show that $\gamma$ intersects each $\Sigma_v$. By causality $\mathbf{g}(\gamma'(v),\gamma'(v)))\le0$, we have 
$$h_{ij}((x^i)'+X^i)((x^j)'+X^j)\le N^2.$$

Consider the distance function $\rho=d(\,\cdot\,,S)$ which extends to $\mathbf{E}$ by stationarity. Note that on $S$, $\mathbf{T}|_S\in TS$,
$$N=-\mathbf{g}(\mathbf{T},n)=0, \quad X(\rho)=\mathbf{T}(\rho)=0.$$
Smoothness up to $S$  gives
$$
 N=O(\rho),\quad X(\rho)=O(\rho),
$$
and we have $|\rho'+X(\rho)|\le |\D \rho(x'+X)|\le N$, and then
\begin{equation}\label{eq:causal-boundary-distance}
 \left|\frac{\D\rho}{\D v}\right|\le C\rho,
 \end{equation}
By Gronwall's inequality, 
$$\rho(v)\ge\rho(v_0)e^{-C|v-v_0|}$$
 as long as the curve remains close to $S$. This means that $\gamma$ cannot reach $S$ at finite $v$. The estimate \eqref{eq:complete-maximal-lapse-bound} gives, near infinity
$$|r'|\le|\D r|(N+|X|)\le C.$$
 This excludes escape to infinity in finite $v$. 
 
 Now suppose that $\gamma$ exists in $[v_0,v_*)$, the above estimates show that the spatial projection $(x^i(v))$ of $\gamma(v)$ in $[v_0,v_*)$ lie in a compact region $K$ of $\Sigma^\circ=\Sigma_0^\circ$.   By causality $|x'|$ is bounded, so as $v$ approaching  $v_1$ from below, $\gamma(v)$ is Cauchy and has a limit in $K$. By reversing the orientation, we conclude that $\gamma$ extends to $v\in(-\infty,+\infty)$ and intersect each $\Sigma_v$.\end{proof}

Fix any sufficiently small $q>0$. we will study the intersection of $\partial I^-_{\mathbf{g}_q}(C_t)$ and the $q$-maximal slices inside the cylinder $\mathcal{C}$. In this paper, all causal relations  are understood relative to $\mathbf{E}$. First we have
 
\begin{lemma}\label{lem:finite-generation}
For every $t\in\mathbb{R}$, every $p\in\partial I^-_{\mathbf{g}_q}(C_t)$ inside $\mathcal{C}$ with $v_q(p)<\min_{C_t}v_q$ is joined to $C_t$ by a future null geodesic in $\partial I^-_{\mathbf{g}_q}(C_t)$. Its final tangent is $L_{q,t}$ after normalization, and no focal point occurs strictly between its endpoints.
\end{lemma}

\begin{proof}
We suppress $t$ in this proof.  By global hyperbolicity and compactness of $C$ we have a causal curve from $p$ to some $c\in C$. Since $p\notin I^-_{\mathbf{g}_q}(C)$, this curve must be a null geodesic on the causal boundary without focal point in its interior. The tangent  at $C$ is orthogonal to $C$. 
 
 If the tangent is not $L_q$, then just before its intersection with $C$, the null geodesic lies outside $\mathcal{C}$.  Take its last intersection with $\mathcal{C}$, at which the temporal function $t<t(C)$ since this last segment lies outside $\mathcal{C}$ and $t$ is increasing on it. The intersection lies in $I^-_{\mathbf{g}_q}(C)$, a contradiction. 
\end{proof}

Fix $\Sigma=\Sigma_0$. All points in $\Sigma^\circ=\Sigma\backslash S$ can be causally connected $C_t$ for sufficiently large $t$. In fact, choose a smooth future-directed $\mathbf g$-timelike curve $\gamma:[0,1]\to\mathbf E$ from the given
point to the asymptotic end, outside $\mathcal{C}$. As in \eqref{eq:timelike-gq},  $\gamma_q$ defined as before is timelike and future directed.  Its spatial path crosses
$\mathcal{C}$, so $\gamma_q$ meets some $C_a$. Following
$\mathbf T$ on the cylinder then reaches $C_t$ for every $t>a$.
This proves the claim.

Below we will suppress $q$.  We will show that the portion of $\Sigma\cap I^-(C_t)$  inside $\mathcal{C}$ can be arbitrarily close to $S$ for sufficiently large $t$. 
\begin{lemma}
For every sufficiently small $\eta>0$, there exists $t_\eta$
such that the following holds for every $t\ge t_\eta$.
Let $K$ be the part of $\Sigma$ inside $\mathcal C$
lying outside $I^-(C_t)$, and set
$$
D=\partial_{\Sigma^\circ}(K\cap\Sigma^\circ).
$$
Then $K$ is compact and contains a neighborhood of $S$ in
$\Sigma$. Moreover, for some $0<\eta_t<\eta$, $D$ is a
compact subset of $\{\eta_t<\rho<\eta\}$ and coincides
with the portion of $\Sigma\cap\partial I^-(C_t)$
inside $\mathcal C$.
\end{lemma}

\begin{proof}
Set $S_\eta=\{x\in\Sigma:\rho(x,S)=\eta\}$.
By the preceding argument, every point of $\Sigma^\circ$
belongs to $I^-(C_t)$ for all sufficiently large $t$.
These chronological pasts are open and increase with $t$,
since $\mathbf T$ is future timelike on $\mathcal C$.
By compactness, we may therefore choose $t_\eta$ so that
the entire region of $\Sigma$ between $S_\eta$ and
$\mathcal C$ is contained in $I^-(C_t)$ for $t\ge t_\eta$.
Increasing $t_\eta$ if necessary also ensures
$\min_{C_t}v>0$.

For fixed $t$, since $v=0$ on $\Sigma$, the total increase in $v$ along
any future-directed causal curve from $\Sigma$ to $C_t$
is at most $\max_{C_t}v$. The estimate \eqref{eq:causal-boundary-distance} and
Gronwall's inequality therefore give a positive lower
bound for its initial distance from $S$.
We can choose
$0<\eta_t<\eta$ such that
$$
\{x\in\Sigma:\rho(x,S)\le\eta_t\}
\cap I^-(C)=\varnothing.
$$
It follows that $K$ contains a neighborhood of $S$,
and that $D$ is compact and lies in
$\{\eta_t\le\rho<\eta\}$.

By definition and standard topology, $D=\partial(\Sigma^\circ\cap K\cap (I^-(C_t))^c) \subset\Sigma^\circ\cap\partial I^-(C_t)$.
Conversely, let $p\in\Sigma^\circ\cap\partial I^-(C_t)$
lie inside $\mathcal C$.
Lemma~\ref{lem:finite-generation} gives a future null
geodesic from $p$ to $C_t$.
Projecting points on a short initial segment of this
geodesic back to $\Sigma$ along a local future timelike
flow (say the gradient of $v$) gives points of $\Sigma\cap I^-(C_t)$ approaching $p$.
Since $p\notin I^-(C_t)$, we have $p\in D$.
\end{proof}

We now fix these choices and suppress $t$.
As $\eta\downarrow0$, we may choose $t\uparrow\infty$
and $\eta_t\downarrow0$.

Denote $A$ be the subset of $C$ so that the past null geodesics from $A$ reach $D$. By Lemma \ref{lem:finite-generation} every point in $D$ can be reached by some past null geodesic from $A$ without focal points in its interior. Since $D$ is closed and then $A$ is also closed by continuity dependence of geodesic equation.  The null geodesics from $A$ to $D$ do not intersect before $D$ (possibly intersect in $D$).   We define a map 
$$F: A\to D$$
by sending points in $A$ along the past null geodesics to $D$. We will choose a smooth approximation of $D$ in $\Sigma$. 

\begin{lemma}\label{lem:smooth-approximation}
For every $\epsilon>0$, there is a smooth compact region $\Omega^\epsilon\subset \Sigma$  such that 
\begin{equation*}
 \begin{split}
 \partial\Omega^\epsilon&=D^\epsilon\sqcup S,\\
 |D^\epsilon|&\le|C|+\epsilon.
 \end{split}
\end{equation*}
Here $D^\epsilon$ is smooth, closed and possibly disconnected. It lies between $S$ and $S_\eta$. \end{lemma}

\begin{proof}
Let $\Psi_z$ be the local flow of the future timelike field $V=Nn=T-X$, with $V(v)=1$. Take the annulus $\mathcal{R}=\{0\le\rho\le \eta\}$ between   $S_\eta$ and $S$. For sufficiently small $\delta>0$, $\Psi_z(\mathcal{R})\supset D_{z}=\Sigma_{z}\cap\partial I^-(C)\cap$ \{\text{inside} $\mathcal{C}$\} for $|z|\le 2\delta$.  On $\mathcal{R}$ we define a function 
$$ u(x)=\sup\bigl(\{-2\delta\}\cup
       \{z\in[-2\delta,2\delta]:\Psi_z(x)\in I^-(C)\}\bigr),$$
       which is the height function of $\partial I^-(C)$ whose values are truncated between $[-2\delta,2\delta]$.  $\delta$ can be chosen small enough so that $u$ takes value $-2\delta$ near $S$ and $2\delta$ near $S_\eta$. Such $u$ is a Lipschitz function on $\mathcal{R}$.  Note that by definition 
       $$\Psi_{-z}(D_z)=\{u=z\}, \quad |z|<2\delta,$$
  we have, for $z\in (-\delta, -\frac{\delta}{2})$, 
  $$\mathcal{H}^2\{u=z\}=\mathcal{H}^2(\Psi_{-z}(D_{z}))\le (1+o_{\delta}(1))\int_{A_z}|J(F_z)|\D\mu_C\le(1+o_{\delta}(1))\int_{A}|J(F)|\D\mu_C $$
  where $\mathcal{H}^2$ is the $2$-dimensional Hausdorff measure, $A_z$ is the subset of $C$ so that the past null geodesics from $A$ reach $D_z$ and $F_z:A\to D_z$ is the corresponding map. For $z<0$ we must have $A_z\subset A$. Also, since $\theta_L(C)>0$,  by Raychaudhuri equation we have $\theta>0$ along the null geodesics without focal points, and in particular $|J(F_z)|\le  |J(F)|\le 1$. By coarea formula, 
  $$\int_{u\in(-\delta,-\frac{\delta}{2})}|\nabla u|\D V=\int_{-\delta}^{-\frac{\delta}{2}} \mathcal{H}^2(\{u=z\})\D z\le  (1+o_{\delta}(1))\frac{\delta}{2} \int_{A}|J(F)|\D\mu_C.$$
  Find $u_i\in C^\infty$ so that $u_i\to u$ in $C^0\cap W^{1,1}$. Then for any $\epsilon>0$, we can find some $\delta$ and $i$ so that (by enlarging $(-\delta,-\frac{\delta}{2})$ a little)
  $$\frac{2}{\delta}\int_{-\delta}^{-\frac{\delta}{2}}\mathcal{H}^2(u_i=z)\D z=\frac{2}{\delta}\int_{u_i\in(-\delta,-\frac{\delta}{2})}|\nabla u_i|\D V\le  \int_{A}|J(F)|\D\mu_C+\frac{\epsilon}{2},$$
  by Sard theorem, we can find some regular value $b\in (-\delta,-\frac{\delta}{2})$ of $u_i$ so that 
  $$|u_i^{-1}(b)|\le \int_{A}|J(F)|\D\mu_C+\epsilon\le |A|+\epsilon\le |C|+\epsilon.$$
  $D^\epsilon=u_i^{-1}(b)$ is then a smooth closed surface. Put $\Omega^\epsilon=\{u_i\le b\}$, the lemma is proved.   
  \end{proof}
  \begin{proof}[Proof of Theorem \ref{thm:S<=4pi}]
  We integrate the vector field $\nabla\rho$ in the region $\Omega^\epsilon$.  Let $\nu$ be the outward unit normal at the boundary, we have
  $$|S|=\int_{D^\epsilon} h(\nabla\rho,\nu)\D\mu -\int_{\Omega^\epsilon}\Delta\rho\D V\le |D^\epsilon| +O(\eta)\le |C|+\epsilon+O(\eta).$$
 Letting $\epsilon\downarrow0$ and then $\eta\downarrow0$ gives $|S|_{\mathbf g_q}\le |C_q|_{\mathbf g_q}$. Finally let $q\downarrow0$ and use \eqref{eq:sharp-C-area-bound} and smooth convergence on $S$ to obtain $|S|_{\mathbf g_0}\le4\pi$. 
  \end{proof}

  The following theorem concerns the reverse inequality.
  \begin{theorem}\label{thm:S>=4pi}
$|S|_{\mathbf{g}_q}\ge 4\pi(1-q^2)$ for all sufficiently small $q>0$. 
\end{theorem}

\begin{proof}

This lower bound is obtained by the  spacetime stability of
the bifurcation sphere $S$. For fixed $q>0$, let $n_q$ be the future unit
normal of $\Sigma$ and $\nu_q$ the outward unit normal of $S$ in
$\Sigma$. Put
$$
 L_q=n_q+\nu_q,\quad
 \underline L_q=\frac12(n_q-\nu_q),\quad
 \mathbf g_q(L_q,\underline L_q)=-1,
 \quad
 (\zeta_q)_A=-\mathbf g_q(\underline L_q,
                         \nabla^{(q)}_{e_A}L_q).
$$
We define the electric and magnetic flux densities $e_q,b_q:S\to\mathbb R$ by
$$
 ({}^{\star_{\mathbf g_q}}\mathbf F_q)|_{TS}=e_q\,\D \mu_q,\quad
 \mathbf F_q|_{TS}=b_q\,\D \mu_q
$$
where $\mu_q$ is the area element of $S$ in $\mathbf{g}_q$. The flux identities \eqref{eq:charge} therefore read
$$
 \int_S e_q\,\D A_q=4\pi\sqrt{1-q^2},\qquad
 \int_S b_q\,\D A_q=0.
$$
Let $\nablas^{(q)}$, $\Deltas^{(q)}$, $\divs^{(q)}$ and $K_q$
denote the connection, Laplacian, divergence and Gauss curvature of
$\mathbf g_q|_{TS}$. The variation of outgoing expansion (relative to $L_q$) under displacement
$\varphi\nu_q$ is
\begin{equation}\label{eq:MOTS-operator}
 \mathscr L_q\varphi
 =-\Deltas^{(q)}\varphi+2\zeta_q\cdot \nablas^{(q)}\varphi+\bigl(K_q+\divs^{(q)}\zeta_q-|\zeta_q|^2                   -e_q^2-b_q^2\bigr)\varphi.
\end{equation}
Here the Einstein--Maxwell equations give
$\operatorname{Ric}_{\mathbf g_q}(L_q,\underline L_q)
=e_q^2+b_q^2$. 
See \cite{AnderssonMarsSimon2008,DainJaramilloReiris2012}.

We will show that the principle eigenvalue $\lambda_1$ of $\mathscr L_q$ is nonnegative, that is, $S$ is a stable MOTS.  If so, by area-charge inequality (see \cite{DainJaramilloReiris2012}), we will have
$$
|S|_{\mathbf g_q}\ge \frac{1}{4\pi}\left(\int_Se_q\,\D \mu_q\right)^2=4\pi(1-q^2)
$$
and the proof is completed. This inequality can be deduced by plugging in $\varphi_q>0$ being the eigenfunction of $\lambda_1$ and Gauss--Bonnet.

 Now suppose that $\lambda_1<0$, and let $\varphi>0$ be its smooth eigenfunction. The outward
normal graphs
$$
 S_\epsilon=\{\exp_x^{h_q}(\epsilon\varphi(\vartheta)\nu_q):\vartheta\in S\}
$$
then have strictly negative outgoing expansion for all sufficiently
small $\epsilon>0$. Choose one such surface then we will have
$$
 \theta_{n_q+\nu_q}(S_\epsilon)\le-\delta
$$
for some $\delta>0$.  As shown in \eqref{eq:timelike-gq}, for each point on $S_\epsilon$, there is a timelike curve from this point to the timelike cylinder $\mathcal{C}=\bigcup_t \varphi^{\mathbf T}_t(C)$, with the value of $v_q$ at the intersection being $B$. We then take a minimal $t_{q,\min}$ so that $C_t\cap J^+(S_\epsilon)\ne\varnothing$. $J^+(S_\epsilon)$ is closed by global hyperbolicity, there is some point $p'\in J^+(S_\epsilon)\cap C_{t_{q,\min}}$.  Then there is some point $p\in S_\epsilon$ so that $p'$ is connected to $p$ by a null geodesic $\gamma$ lying in $\partial J^+(S_\epsilon)$ without focal points in its interior and orthogonal to $S_\epsilon$ at $p$ and orthogonal to $C_{t_{q,\min}}$ at $p'$.  The tangent of $\gamma$ at $p$ is positively proportional to $n_q+\nu_q$, otherwise,  $\gamma$ will reach $\mathcal{H}^+$ and escape from  $\mathbf{E}$. 
 By replacing the above $C$ by an arbitrarily faraway coordinate sphere, which enters the asymptotic region of $\mathbf{g}_q$, the same argument applies and the affine length of $\gamma$ can be made larger than $\frac{2}{\delta}$ by asymptotic flatness, if we parametrize $\gamma$ so that $\gamma'(0)=n_q+\nu_q$, a contradiction by Raychaudhuri equation.


\end{proof}

\begin{corollary}\label{cor:Se0b0}
The bifurcation sphere satisfies
$$ |S|_{\mathbf g_0}=4\pi,\quad e_0=1,\quad b_0=0.$$
\end{corollary}

\begin{proof}
Combining Theorems \ref{thm:S<=4pi} and \ref{thm:S>=4pi} and let $q\downarrow0$, we will obtain $|S|_{\mathbf{g}_0}=4\pi$.  Again from \eqref{eq:MOTS-operator} we have, for sufficiently small $q>0$, (see also \cite{DainJaramilloReiris2012})
$$\int_S(e_q^2+b_q^2)\D\mu_q\le 4\pi.$$
 Passing to the limit we have
$$
 \int_S(e_0^2+b_0^2)\,\D \mu_0\le4\pi.
$$
Together with the area of $S$, we have,
$$0\le \int_S ((e_0-1)^2+b_0^2)\D\mu_0\le \int_S\bigl((e_0^2+b_0^2)-2e_0+1\bigr)\le 4\pi-8\pi+4\pi=0.$$
Thus the left hand side equal zero and $e_0=1$, $b_0=0$ pointwisely. 
\end{proof}

\section{Recovery of the Ionescu--Klainerman conditions}
\label{sec:area-stability-ik}

We now use Corollary~\ref{cor:Se0b0} to verify the two technical
assumptions of Ionescu--Klainerman \cite{IonescuKlainerman2009}.
The local horizon field $\mathbf{K}$ has boost rate $\kappa>0$.
We verify the scalar inequality at the zero $p\in S$ of $\mathbf T$
fixed in Section~\ref{sec:harrison}.

\begin{proposition}\label{prop:area-stability-ik}
The original vacuum metric satisfies
\begin{equation}\label{eq:area-stability-ik-conditions}
 -4\mathcal{F}_{ab}\mathcal{F}^{ab}=(1-\sigma)^4
 \quad\hbox{on }S,\qquad
 \operatorname{Re}\frac1{1-\sigma}>\frac12
 \quad\hbox{at }p,
\end{equation}
where indices on $\mathcal{F}$ are raised using the original metric
$\mathbf g$.
\end{proposition}

\begin{proof}
By Corollary~\ref{cor:Se0b0}, $e_0=1$ and $b_0=0$ on $S$.
The $\mathbf{K}$-flow fixes $S$, acts as a nontrivial boost on its normal
plane, and preserves $\mathbf F_0$. Thus the mixed normal--tangential
components of $\mathbf F_0$ vanish.  In an oriented $\mathbf{g}_0$-orthonormal frame
$(n=e_0, \nu=e_1,e_2, e_3)$ for $\mathbf g_0$, with $e_0$ future timelike and $e_1$
outward spacelike, we have, on $S$, 
$$(\mathbf{F}_0)_{23}= ({}^{\star_{\mathbf g_q}}\mathbf F_0)_{01}=b_0=0, \quad ({}^{\star_{\mathbf g_q}}\mathbf F_0)_{23}=-(\mathbf{F}_0)_{01}=e_0=1,$$
and the other components vanish. By \eqref{eq:transformed-two-form}, we have
\begin{equation}\label{eq:Harrisontran-g0}
\mathcal F
=-\frac{1-\sigma}{1-\bar\sigma}
(A_0^{-1})^*
\left(\mathbf F_0+\mathrm i{}^{\star_{\mathbf g_0}}\mathbf F_0\right).
\end{equation}
By \eqref{eq:Dq-definition} and \eqref{eq:harrison-metric} we have
$$
 \mathbf g_0=\frac{|1-\sigma|^2}{4}A_0^*\mathbf g.
$$
Therefore, contracting $\mathcal F$  using the original metric $\mathbf{g}$, and $\mathbf{F}_0+\mathrm i{}^{\star_{\mathbf g_0}}\mathbf F_0$ by $\mathbf{g}_0$, we have on $S$
$$\mathcal F_{ab}\mathcal F^{ab}=\left(\frac{1-\sigma}{1-\bar\sigma}\right)^2\frac{|1-\sigma|^4}{16}\left( \mathbf F_0+\mathrm i{}^{\star_{\mathbf g_0}}\mathbf F_0\right)_{ab}\left( \mathbf F_0+\mathrm i{}^{\star_{\mathbf g_0}}\mathbf F_0\right)^{ab}=-\frac{1}{4}(1-\sigma)^4.$$
This is the first condition. 

For the second, we work at $p$. Since $\mathbf T(p)=0$,
we have $A_0(p)=\mathrm{id}$ and $f(p)=0$. Thus
$\mathbf g_0(p)=\frac{d_0}{4}\mathbf g(p)$ and $\bar\sigma(p)=-\sigma(p)$,
so by \eqref{eq:Harrisontran-g0} at $p$, 
$$
\mathcal F
=-\frac{1-\sigma}{1+\sigma}
\left(\mathbf F_0+\mathrm i{}^{\star_{\mathbf g_0}}\mathbf F_0\right).
$$
Recalling that $\nabla_a\mathbf{T}_b=\mathrm{Re\mathcal F}_{ab}$,  and $\left(\mathrm{Re}\left(\mathbf F_0+\mathrm i{}^{\star_{\mathbf g_0}}\mathbf F_0\right)\right)_{01}=-1$, we have
$$
 \mathbf{g}(\nabla_{n}\mathbf{T},\nu)(p)
 =\operatorname{Re}\frac{1-\sigma(p)}{1+\sigma(p)}
 =2\operatorname{Re}\frac1{1-\sigma(p)}-1.
$$
We only need to show that the $\mathbf{g}(\nabla_{n}\mathbf{T},\nu)>0$ at $p$. 

Because $\mathbf T(p)=0$, we have $\nabla^{(0)}_n\mathbf T(p)=\nabla_n\mathbf T(p)$ and
$$
\mathbf g(\nabla_n\mathbf T,\nu)(p)=\mathbf g(\nabla_n^{(0)}\mathbf T,\nu)(p)
 =\frac{4}{d_0}\mathbf g_0(\nabla^{(0)}_n\mathbf T,\nu)(p).
$$
Finally, recall that $\mathbf K=\mathbf T+\Omega_H \mathbf Z$ so 
$$\nabla^{(0)}_n\mathbf K=\nabla^{(0)}_n\mathbf T+\Omega_H \nabla^{(0)}_n\mathbf Z.$$
 By averaging over the orbit of $\mathbf Z$, we obtain a $\mathbf Z$-invariant future unit timelike normal $\bar{n}$, with $[\bar{n},\mathbf Z]=0$. Then
$$\nabla^{(0)}_{\bar{n}}\mathbf Z=\nabla^{(0)}_{\mathbf Z} \bar{n}=0$$
at $p$ by $\mathbf{Z}(p)=0$. We choose a outward spacelike normal $\bar{\nu}$ orthogonal to $\bar{n}$ and simply replace the orthonormal frame by $(\bar{n}, \bar{\nu})$, then $$2\operatorname{Re}\frac1{1-\sigma(p)}-1=\mathbf g(\nabla_{\bar{n}}\mathbf T,\bar{\nu})(p)=\frac{4}{d_0}\mathbf g_0(\nabla^{(0)}_{\bar n}\mathbf K,\bar{\nu})(p)=\frac{4\kappa}{d_0}>0$$
and the proof is completed.

\end{proof}

\end{document}